\documentclass[11pt]{article}

\usepackage{epsfig,graphics}

\usepackage{amssymb}

\newcommand{\bPi}{\mbox{\boldmath $\Pi$}}
\newcommand{\bPsi}{\mbox{\boldmath $\Psi$}}
\newcommand{\bP}{\mbox{\boldmath $P$}}

\newcommand{\beqn}{\begin{eqnarray}}
\newcommand{\eeqn}{\end{eqnarray}}
\newcommand{\be}{\begin{equation}}
\newcommand{\ee}{\end{equation}}
\newcommand{\ba}{\begin{array}}
\newcommand{\ea}{\end{array}}
\newcommand{\R}{{\rm\bf R}}
\newcommand{\C}{{\rm\bf C}}
\newcommand{\pa}{\partial}

\newcommand{\re}{\ref}
\newcommand{\ci}{\cite}
\newcommand{\la}{\label}
\newcommand{\bfr}{\begin{flushright}}
\newcommand{\efr}{\end{flushright}}
\newcommand{\bfl}{\begin{flushleft}}
\newcommand{\efl}{\end{flushleft}}
\newcommand{\fr}{\frac}
\newcommand{\ov}{\overline}
\newcommand{\ti}{\tilde}
\newcommand{\st}{\stackrel}

\newcommand{\si}{\sigma} 
\newcommand{\al}{\alpha}

\newcommand{\ds}{\displaystyle}
\newcommand{\ts}{\textstyle}

\newcommand{\cE}{{\cal E}} \newcommand{\cF}{{\cal F}}
\newcommand{\cH}{{\cal H}}

\newcommand{\cO}{{\cal O}}
\newcommand{\cS}{{\cal S}}

\newcommand{\bI}{{\bf I}}
 \newcommand{\cT}{{\cal T}}
\newcommand{\cZ}{{\cal Z}}

\newcommand{\5}{{\hspace{0.5mm}}}
\newcommand{\ve}{\varepsilon}
\newcommand{\vp}{\varphi}
\newcommand{\we}{\wedge}
\newcommand{\de}{\delta}\newcommand{\De}{\Delta}

 \newcommand{\ga}{\gamma}
\newcommand{\om}{\omega}
\newcommand{\Om}{\Omega}

\newcommand{\na}{\nabla}

\newcommand{\lam}{\lambda} \newcommand{\ka}{\varkappa}
\newcommand{\Lam}{\Lambda}
\newcommand{\co}{{\rm const}}
\newcommand{\supp}{{\rm supp\5}}

\newcommand{\Br}{|\kern-.25em|\kern-.25em|}
\newcommand{\brr}{{|\kern-.15em|\kern-.15em|\kern-.15em}\,}
\newcommand{\ddd}{\st{.\kern-.07em.\kern-.07em.}}
\def\N{{\rm I\kern-.1567em N}}                              % Doppel-N
\def\R{{\rm I\kern-.1567em R}}                              % Doppel R
\def\C{{\rm C\kern-4.7pt                                    % Doppel C
\vrule height 7.7pt width 0.4pt depth -0.5pt \phantom {.}}}
\def\Z  {{\sf Z\kern-4.5pt Z}}                                % Doppel Z
\def\Re {{\rm Re\, }}                                       % Re
\def\Im {{\rm Im\,}}                                        % Im

\begin{document}

\renewcommand{\theequation}{\thesection.\arabic{equation}}
\newtheorem{theorem}{Theorem}[section]
\renewcommand{\thetheorem}{\arabic{section}.\arabic{theorem}}
\newtheorem{definition}[theorem]{Definition}
\newtheorem{deflem}[theorem]{Definition and Lemma}
\newtheorem{lemma}[theorem]{Lemma}
\newtheorem{example}[theorem]{Example}
\newtheorem{remark}[theorem]{Remark}
\newtheorem{remarks}[theorem]{Remarks}
\newtheorem{cor}[theorem]{Corollary}
\newtheorem{pro}[theorem]{Proposition}

\newcommand{\bd}{\begin{definition}}
 \newcommand{\ed}{\end{definition}}
\newcommand{\bt}{\begin{theorem}}
 \newcommand{\et}{\end{theorem}}
\newcommand{\bqt}{\begin{qtheorem}}
 \newcommand{\eqt}{\end{qtheorem}}

\newcommand{\bp}{\begin{pro}}
 \newcommand{\ep}{\end{pro}}

\newcommand{\bl}{\begin{lemma}}
 \newcommand{\el}{\end{lemma}}
\newcommand{\bc}{\begin{cor}}
 \newcommand{\ec}{\end{cor}}

\newcommand{\bex}{\begin{example}}
 \newcommand{\eex}{\end{example}}
\newcommand{\bexs}{\begin{examples}}
 \newcommand{\eexs}{\end{examples}}

\newcommand{\bexe}{\begin{exercice}}
 \newcommand{\eexe}{\end{exercice}}

\newcommand{\br}{\begin{remark} }
 \newcommand{\er}{\end{remark}}
\newcommand{\brs}{\begin{remarks}}
 \newcommand{\ers}{\end{remarks}}

\newcommand{\ft}{\footnote}

\newcommand{\pru}{{\bf Proof~~}}

\begin{titlepage}

\begin{center}
{\Large\bf
Scattering Asymptotics for a Charged Particle\\[2ex]
Coupled to the  Maxwell Field}\\
\vspace{2cm} {\large Valery Imaykin} \footnote{Supported partly by DFG grant 436 RUS
113/929/0-1, FWF Project P19138-N13, RFBR grant 07-01-00018a, and RFBR-DFG grant
08-01-91950-NNIOa. }
\medskip\\
{\it Zentrum Mathematik, TU M\"{u}nchen\\ Boltzmannstr. 3,
Garching, 85747 Germany}\\
email: imaikin@ma.tum.de\bigskip\\
{\large Alexander Komech}$^{1,}$\footnote{Supported partly by Alexander von Humboldt
Research Award.}
\medskip\\
{\it Faculty of Mathematics of
Vienna University\\
Nordbergstrasse 15, 1090 Vienna, Austria
}\\
email: alexander.komech@univie.ac.at\bigskip\\
{\large Herbert Spohn}\medskip\\
{\it Zentrum Mathematik, TU M\"{u}nchen\\ Boltzmannstr. 3,
Garching, 85747 Germany}\\
email: spohn@ma.tum.de\bigskip\\

\end{center}

\vspace{2cm}

\begin{abstract}

We establish long time soliton asymptotics for the nonlinear system of Maxwell
equations coupled to a charged particle. The coupled system has a six dimensional
manifold of soliton solutions. We show that in the long time approximation, any
solution, with an initial state close to the solitary manifold, is a sum of a soliton
and a dispersive wave which is a solution of the free Maxwell equations. It is
assumed that the charge density satisfies the Wiener condition. The proof further
develops the general strategy based on the symplectic projection in Hilbert space
onto the solitary manifold, modulation equations for the parameters of the
projection, and decay of the transversal component.
\end{abstract}
\end{titlepage}

%%%%%%%%%%%%%%%%%%%%%%%%%%%%%%%%%%%%%%%%%%%%%%%%%%%%%%%%%%%%%%%%%%%%%%%%%%%

\setcounter{equation}{0}

\section{Introduction}

Our paper deals with an old and important problem of mathematical physics, namely,
the problem of particle-field interaction. The equations of motion of a charged
particle in external electromagnetic fields was introduced by Lorentz in 1892
\ci{Lor}, though for the first time it was written down by Maxwell in one of his
investigations in the 1860ies. On the other hand, formulas for the electromagnetic
field generated by a moving charge were obtained by Li\'{e}nard and Wiechert
independently in 1898, resp. in 1900. Thus the problem of the interaction of a charge
with its self-generated field arises. The Li\'{e}nard-Wiechert potentials imply that
the field generated by an accelerated charge transports energy to infinity, hence the
acceleration should tend to zero as $t \to \infty$. This {\it radiative decay} is
known since Abraham   \ci{Abr} (1905) and is claimed in most of manuals on
electrodynamics. However, it was proven only fairly recently in \ci{KSK,KS} for the
model of the scalar field coupled to extended charge, and in \ci{IKMml,KS00} for the
Maxwell field coupled to extended charge, as introduced by Abraham. The corresponding
scalar or Maxwell fields converge to the static solutions  in the models with an
external confining potentials \ci{KSK,KS00}, or to the solitons
 (travelling wave solutions)
in the translation invariant models \ci{IKMml,KS}.
Here we refine the asymptotics \ci{IKMml} for the Maxwell-Lorentz equations
identifying the outgoing dispersive wave and the rate of the
convergence for initial states close to a soliton.

It is convenient to write the equations of motion in Hamiltonian form. The dynamical
variables come then in canonically conjugate pairs. They are the position, $q$, of
the particle, together with its momentum $P$, and the transverse vector potential,
$A$, together with the transverse electric  field $E$. We refer to \ci[Chapter
13]{Sp} for details. In these variables the Hamiltonian function reads \be\la{hamil}
{\cal H}(E,A,q,P)=\fr 12\langle E,E\rangle+\fr 12\langle\na A,\na A\rangle+\left[
1+(P-A_\rho(q))^2\right]^{1/2}. \ee on the subspace defined by the transversality
conditions \be\la{csh} \na\cdot E(x)=0,\quad\na\cdot A(x)= 0. \ee Here
$\langle\cdot,\cdot\rangle$ denotes the $L^2$-scalar product and, written in
components,
$$
\langle E,E\rangle=\sum\limits_{j=1}^3\langle E_j,E_j\rangle, \quad \langle\na A, \na
A\rangle=\sum\limits_{i,j=1}^3\langle \na_i A_j ,\na_i A_j\rangle.
$$
The function $\rho$ is the charge density and $A_\rho$ the convolution
$$
A_\rho(x) =\int\,d^3x'\rho(x'-x)A(x').
$$
 The
canonical equations of motion follow then as \be\la{fsh} \dot E(x,t)=-\Delta
A(x,t)-\Pi_s(\rho(x-q(t))\dot q(t)),\,\,\, \dot A(x,t)=-E(x,t),
\ee
\be\la{ph} \dot
q(t)=\fr{P(t)-A_\rho(q(t))}{\left
[1+\big(P(t)-A_\rho(q(t))\big)^2\right]^{1/2}},~~~~~~
 \dot P(t)=[\na(\dot q(t)\cdot A)]_\rho(q(t),t)
\ee
with $t\in\R$;
$x,q,P\in\R^3$.  Here and below all derivatives are understood in the sense
of distributions. The operator $\hat\Pi_s$ is the projection onto the space of
solenoidal (divergence-free) vector fields, which in Fourier space reads
$$
\hat\Pi_s(k)\, a=a-\fr{a\cdot k}{k^2}k.
$$
 It is easily checked that the transversality condition is preserved in time. We
use units such that the velocity of light $c=1$, $\ve_0=1$, and the mechanical mass
of the charge $m=1$.

%We note that, because of translational invariance, the total momentum \be\la{mom-ham}
%{\cal P}(E,A,q,P)=P+\int \,E(x)\we(\na\we A(x))d^3x \ee is conserved along
%sufficiently smooth trajectories of (\re{fsh})-(\re{csh}).

Let us write the system (\re{csh})-(\re{ph}) as \be\la{HPDE} \dot
Y(t)=F(Y(t)),~~~~~t\in\R, \ee where $Y(t)=(E(x,t),A(x,t),q(t),P(t))$
and the phase
space is defined through ${\cal H}<\infty$. Below we always deal with
column vectors
but often write them as row vectors. The system (\re{csh})-(\re{ph})
admits special
solutions where the charge travels with constant velocity. In analogy
with travelling
solutions of nonlinear wave equation we call them {\it solitons}.
Explicitly they are
given by
\be\la{sosol}
Y_{a,v}(t)=(E_{v}(x-vt-a),A_v(x-vt-a),vt+a,P_v),~~~~P_v=p_v
+\langle\rho,A_v\rangle
\ee
for all $a,v\in\R^3$ with $|v|<1$, where $E_{v}=\Pi_sE^v$,
$A_{v}=\Pi_sA^v$, and $E^v$, $A^v$, $p_v$ are given by
\be\la{solformula}
\left.\ba{l}
E^v(x)=-\nabla\phi_v(x)+v\cdot\nabla~A^v(x),\,\,\,A^v(x) =v\phi_v(x),\\
\phi_v(x)=\fr{\ts\ga}{\ts 4\pi}\ds\int \fr{\ts\rho(y)d^3y}{\ts|\ga(y-x)_\|+(y-x)_\bot |},\,\,\,
p_v=\ga v.
\ea\right|
\ee
Here $\ga=1/\sqrt{1-v^2}$ and $x=x_\|+x_\bot$ with $x_\|$ the component parallel and
$x_\bot$ the component orthogonal to $v$. The formulas (\re{solformula}) follow resolving the stationary equations which read
\be\la{solv}
\left.\ba{l}
E_{v}(x)=v\cdot\na A_{v}(x),~~~~~~~~~~~~~~~~~ v\cdot\na E_{v}(x)=\Delta
A_{v}(x)+\Pi_s(\rho(x)v),\\
v=\fr{\ts P_{v}-\langle\rho,A_{v}\rangle}{\ts\left[1+(P_{v}-\langle
\rho,A_{v}\rangle)^2\right]^{1/2}},~~~~~~~~~0=\ds\int \,\rho(x)v\cdot\na A_{v}(x)d^3x.
\ea\right|
\ee

The states $S_{a,v}=Y_{a,v}(0)$ form the solitary manifold \be\la{soman} {\cal S}=\{
S_{a,v}: a,v\in\R^3, |v|<1 \}. \ee For general initial data one expects that for
large times the solution splits up into two parts: one piece consists of a soliton
with a definite velocity and the second piece are scattered fields escaping to
infinity. In fact, this will be our main result. If the initial data are close to the
solitary manifold, then we will prove that for large $t$ \be\la{Si} (E(x,t),
A(x,t))\sim (E_{v_\pm}(x-v_\pm t-a_\pm), A_{v_\pm}(x-v_\pm
t-a_\pm))+W^0(t)\bPsi_\pm,~~~~~~t\to\pm\infty. \ee Here $W^0(t)$ is the dynamical
group of the free wave equation (Equations (\re{fsh}) with $\rho=0$ and (\re{csh})),
$\bPsi_\pm$ are the corresponding {\it asymptotic scattered fields}, and the
remainder converges to zero {\it in the global energy norm}, i.e. in the norm of the
space ${\cal F}:=H^0_s(\R^3)\oplus \dot H^1_s(\R^3)$, see Section 2. For the particle
trajectory we prove that \be\la{qqi} \dot q(t)\to v_\pm, ~~~~q(t) \sim v_\pm t+a_\pm,
~~~~~~~t\to\pm\infty. \ee The results are established under the following conditions
on the charge distribution: $\rho$ is a real valued function of the Sobolev class
$H^2(\R^3)$, compactly supported, and spherically symmetric, i.e. \be\la{ro}
\rho,\na\rho,\na\na\rho\in L^2({\R}^3),~~~~~~~~\,\,\,\,\,\,\quad\rho (x)=0
\,\,\,\mbox{for}\,\,\,|x|\ge R_\rho, \,~~~~~~~~~~~~~\rho (x)=\rho_1(|x|). \ee An
essential point of our asymptotic analysis is the Wiener condition \be\la{W}
\hat\rho(k)=(2\pi)^{-3/2}\int\limits \, e^{ik x}\rho(x)d^3x\not=0 \mbox{
\,\,\,for\,\,all\,\, }k\in\R^3\setminus\{0\}\,. \ee The Wiener condition was noted
already in the previous works \ci{
%IKMml,IKV05,IKV06,
IKMml,KS,KS00}. It expresses that
all modes of the Maxwell field are coupled to the particle.

There is no restriction on $\int|\rho(x)|d^3x$. However if $\int\rho(x)d^3x\ne0$,
then the soliton fields fields have a slow decay at infinity, namely,
$A_v(x)\sim|x|^{-1}$ and $E_v(x)\sim|x|^{-2}$. With our methods such a decay seems to
be difficult to control and we have to impose the condition of vanishing the momenta
of $\rho$ up to the fourth order: \be\la{neutr} \int\,
x^{\al}\rho(x)d^3x=0,\,\,\,\,|\al|\le4. \ee In particular, the total charge
$\int\rho(x)d^3x$ equals zero (neutrality of the particle). Equivalently, $\hat\rho$
has a fifth order zero at $k=0$; \be\la{zero5}
\hat\rho^{\,(\alpha)}(0)=0,\,\,\,\,|\al|\le4. \ee We believe (\re{neutr}) to be a
technical condition. Physically one expects (\re{Si}) to hold even without imposing
charge neutrality and it is of interest to extend our proof in this direction.

Let us briefly comment on earlier works. The first mathematical investigation is the
contribution of Bambusi and Galgani \ci{BG}. They consider a non-relativistic kinetic
energy for the charge and prove orbital stability of the solitons,
without Wiener condition.
The asymptotics of type (\re{Si}) for the fields alone, without $q,\dot q$ were proved
under the Wiener condition for charged particle coupled to scalar or
Maxwell field with
 a potential in \ci{KSK,KS00} and for the translation invariant
systems without potential in \ci{IKMml,KS}. However, the asymptotics were proved only
in the local energy semi-norms and did not involve the dispersive term.

Full asymptotics (\re{Si}), (\re{qqi}) were established under the weak
coupling condition
$\Vert\rho\Vert_{L^2}\ll1$ for translation invariant Maxwell-Lorentz
system in \ci{IKSm}
and for Maxwell-Lorentz system with a rotating particle in \ci{IKSr}.
In the present paper we establish the full asymptotics (\re{Si}),
(\re{qqi}) without the
weak coupling condition under the Wiener condition (\re{W}).

Long time asymptotics of type (\re{Si}) appears also in nonlinear wave equations,
like the KDV \ci{EH81,FZ93} and the $U(1)$-invariant nonlinear Schr\"odinger equation
\ci{BP1,BP2,BP3,BS, PW92,PW94, SW88, SW90,SW92}. In these equations there are no
particle degrees of freedom and the solitons (\re{sosol}) correspond to the solitary
wave solutions travelling at constant velocity.

%%%%%%%%%%%%%%%%%%%%%%%%%%%  25.09.2010 %%%%%%%%%%%%%%%%%%%%%%%%%

Let us comment on basic peculiarity of our problem. Namely,  the asymptotics
(\re{Si}), (\re{qqi}) mean the asymptotic stability of the solitary manifold $\cS$ in
the dynamics (\re{fsh})-(\re{ph}). However, the dynamics {\it along} the solitary
manifold is unstable, and this is the main difficulty in the proofs. Namely, for two
soliton solutions with close but different velocities $v_1$ and $v_2$ and close
initial positions $q_1^0$ and $q_2^0$ one has
$$
q_1(t)-q_2(t)=q_1^0-q_2^0+(v_1-v_2)t\to\infty\,\,\,{\rm as}\,\,\,t\to\infty.
$$
Moreover, the fields
$(E_1(x,t),A_1(x,t))$ and $(E_2(x,t),A_2(x,t))$ being close at $t=0$
do not remain so
as $t\to\infty$, since they are centered at $q_1(t)$ and $q_2(t)$,
although their
difference remains bounded. The nonlinear instability
corresponds to the fact that
tangent vectors
$\pa_{a_j} S_{a,v}$ and $\pa_{v_j} S_{a,v}$, $j=1,2,3$
to the solitary manifold are the zero eigenvectors and root vectors
for the
generator of the
linearized equation.
Respectively,
the linearized equation
admits linear in
$t$ secular solutions, see (\re{secs}). The existence of these runaway
solutions
prohibits the direct application of the Liapunov strategy and
requires significant modification of the classical stability theory.

Our approach
relies on and further develops the general  strategy introduced in
the cited papers in the context
of the
$U(1)$-invariant Schr\"odinger equation.
The approach uses i) symplectic projection of the dynamics
in the Hilbert phase space onto the
symplectic orthogonal
directions to the solitary manifold to kill the  runaway
 secular solutions, ii) the modulation equations
for the motion along the  solitary manifold, and iii) freezing of the dynamics in the
non-autonomous linearized equation. See more details in Introduction \ci{IKV05} where
the general strategy has been developed for the case of the Klein-Gordon equation.
The
 Maxwell-Lorentz equations
 (\re{fsh})-(\re{ph})
differs significantly from the  Klein-Gordon case because of slow Coulombic decay of
the solitons and presence of the embedded eigenvalue in the continuous spectrum of
the linearized equation (see the comments below).

Developing the general strategy for the Maxwell-Lorentz equations
 (\re{fsh})-(\re{ph}),
we obtain our main result  in Sections 3-9 and Appendix A of the paper. The main
novelty in our case is thorough establishing the appropriate decay of the linearized
dynamics in Sections 10-13 and Appendices B and C:

I. We
do not postulate any spectral properties of the linearized equation,
calculating all the properties from the Wiener condition (\re{W}).
Namely, we show that
 i) the full zero spectral space
of the linearized equation is spanned by the
tangent vectors,
and moreover, ii)
there are no others (nonzero) discrete eigenvalues
(see Lemmas \re{PiomQiom}, \re{sort123}
and Proposition \re{regi}).

II. Using these spectral properties,
we prove
that the linearized equation is stable in
the {\it
symplectic orthogonal complement} to the tangent space $\cT_S$
spanned by the
tangent vectors
$\pa_{a_j} S_{a,v}$ and $\pa_{v_j} S_{a,v}$, $j=1,2,3$.
We exactly
calculate
 in
Lemma \re{sort123}
the corresponding symplectic orthogonality conditions
for initial data of the linearized dynamics.

III. One of main peculiarities of the Maxwell-Lorentz equations is the presence of
embedded eigenvalue $\lam=0$ in the continuous spectrum $\si_c=\R$ of the linearized
equation. This situation never happens in all previous works on the asymptotic
stability of the solitary waves for the  Schr\"odinger and  Klein-Gordon equations.
Thus, the symplectic orthogonality condition is imposed now at the interior point of
the continuous spectrum in contrast to all previous works in the field. Respectively,
the integrand at this point in the spectral representation of the solution is not
smooth even if the  symplectic orthogonality condition holds. Hence, the integration
by parts in this spectral representation as in the case of the  Schr\"odinger and
Klein-Gordon equation, is impossible. For the proof of the decay in this new
situation, we transform the spectral representation in  the proofs of Propositions
\re{r2p2} and \re{r1p1}, and develop
 new
more subtle technique of convolutions.

\medskip

Our paper is organized as follows. In Section 2, we formulate the
main result. In  Section 3, we introduce the symplectic projection
onto the solitary manifold. The linearized equation is defined and
studied in Sections 4-5. In Section 6, we split the dynamics in
two components: along the solitary manifold, and in transversal
directions. In Section 7 we justify the slow motion of the longitudinal
component, and in Section 8
the decay of the transversal
component assuming the corresponding
decay in the linearized dynamics, which is proved in
Sections 10-13. In Section 9 we prove the main
result.
In Appendices A, B, and C we collect routine calculations.
~\medskip\\
{\bf Acknowledgement}. The authors thank V. Buslaev
for numerous lectures on his
results and fruitful discussions, and E. Kopylova for
improvements and corrections.

%%%%%%%%%%%%%%%%%%%%%%%%%%%%%%%%%%%%%%%%%%%%%%%%%%%%%%%%%%%%%%%%%%%%%%%%

\setcounter{equation}{0}

\section{Main Results}

%%%%%%%%%%%%%%%%%%%%%%%%%%%%%%%%%%%%%%%%%%%%%%%%%%%%%%%%%%%%%%%%%%%%%%%%

\subsection{Existence of Dynamics}

Let us introduce a phase space for the system (\re{csh})-(\re{ph}) and state the
existence of dynamics. Set $H^0=L^2(\R^3,\R^3)$, $\dot H^1$ is the closure of
$C_0^{\infty}(\R^3,\R^3)$ with respect to the norm $\Vert A\Vert_1=\Br\na
A\Br=\Vert\na A\Vert_{L^2(\R^3,\R^3)}$. Let $H^0_s$, $\dot H^1_s$ be the subspaces
constituted by solenoidal vector fields, namely the closure in $H^0$, $\dot H^1$
respectively of $C_0^{\infty}$ vector fields with vanishing divergence. Define the
phase space
$$
{\cal E}=H^0_s \oplus \dot H^1_s\oplus \R^3 \oplus
\R^3,\,\,\,Y=(E,A,q,P),\,\,\,\Vert Y\Vert_{\cal E}=\Br E\Br+\Vert
A\Vert_1+|q|+|P|.
$$
Let us define the corresponding space for fields alone:
$$
{\cal F}=H^0_s \oplus \dot H^1_s,\,\,\,\Vert(E,A)\Vert_{\cal F}=\Br E\Br+\Vert A\Vert_1.
$$
We write the Cauchy problem for the system (\re{csh})-(\re{ph}) as
\be\la{1dh}
\dot Y=F(Y(t)),\,\,\,t\in\R;\,\,\,Y(0)=Y^0.
\ee

\begin{pro}\la{ex-ham} {\rm \ci{IKMml}}.
Let (\re{ro}) holds, let $Y^0=(E^0,A^0,q^0,P^0)\in{\cal E}$. Then

\noindent i) There exists a unique solution $Y(t)\in C(\R,{\cal E})$ to the Cauchy
problem (\re{1dh}).

\noindent ii) The energy conserves,
$$
H(Y(t))=H(Y^0),\,\,\,t\in\R.
$$

\noindent iii) The estimate holds,
\be\la{ov-v}
|\dot q(t)|\le\ov v<1,\,\,\,t\in\R.
\ee
\end{pro}

%%%%%%%%%%%%%%%%%%%%%%%%%%%%%%%%%%%%%%%%%%%%
%%%%%%%%%%%%%%%%%%%%%%%%%%%%%%%%%%%%%%%%%%%%

\subsection{The Main Result}

To state our main result we have to introduce the following weighted Sobolev spaces.
Let $H^0_{s,\alpha}$, $H^1_{s,\alpha}$ be the subspaces of $H^0_s$, respectively
$\dot H^1_s$ consisting of all the fields $E$, resp. $A$ with the finite norms
$$
\Vert E\Vert_{0,\alpha}=\Br(1+|x|)^\alpha E\Br,\,\,\,\Vert A\Vert_{1,\alpha}=\Vert A\Vert_{0,\alpha}+\Br(1+|x|)^\alpha\na A\Br.
$$
%Since $\dot H^1_s\subset L^6(\R^3,\R^3)$, we have $\dot H^1_s\subset H^0_{s,\alpha}$,
%$\alpha<-1$.
Let us define
$$
{\cal E}_\alpha=H^0_{s,\alpha+1}\oplus H^1_{s,\alpha}\oplus\R^3\oplus\R^3,\,\,\,
\Vert Y\Vert_\alpha=\Vert E\Vert_{0,\alpha+1}+\Vert A\Vert_{1,\alpha}+|q|+|P|,\,\,\,
Y\in{\cal E}_\alpha.
$$
For the fields we set
$$
{\cal F}_\alpha=H^0_{s,\alpha+1}\oplus H^1_{s,\alpha},\,\,\,
\Vert(E_s,A)\Vert_\alpha=\Vert E\Vert_{0,\alpha+1}+\Vert A\Vert_{1,\alpha}.
$$

\begin{definition}
A soliton state is $S(\si):=(E_v(x-b),A_v(x-b),b,P_v)$, where
 $\si:=(b,v)$ with
 $b,v\in\R^3$ and $|v|<1$.
\end{definition}
Obviously, the soliton solution admits the representation $S(\si(t))$, where
\be\la{sigma}
\si(t)=(b(t),v(t))=(vt+a,v).
\ee
\begin{definition}
A solitary manifold is the set ${\cal S}:=\{S(b,v):b\in \R^3,
|v|<1\}$.
\end{definition}
By (\re{solformula}) and the condition (\re{neutr}) we obtain that
$$
A_v(y)={\cal O}(|y|^{-6}),\,\,\,E_v(y)={\cal O}(|y|^{-7}),\,\,\,|y|\to\infty.
$$
Thus,
$$
E_v\in H^0_{s,\alpha}\,\,\,{\rm for}\,\,\,\alpha<11/2,\,\,\,A_v\in\dot H^1_{s,\alpha}\,\,\,{\rm for}\,\,\,\alpha<9/2
$$
and we have for the soliton states
\be\la{solw}
S(\si)\in{\cal E}_{\alpha},\,\,\,{\rm
for}\,\,\,\alpha<9/2.
\ee
The main result of our paper is the following theorem.
\begin{theorem}\la{main}
Let the condition (\re{ro}), Wiener condition (\re{W}), and the condition
(\re{neutr}) hold, let $\beta=4+\de$, $0<\de<1/2$.
 Suppose that the initial state $Y^0\in{\cal E}_{\beta}$ and  is sufficiently
close to the solitary manifold:
\be\la{close}
Y^0=S_{a_0,v_0}+Z_0,\,\,\,\,d_\beta:=\Vert Z_0\Vert_{\beta}\ll 1.
\ee
Let $Y(t)\in C(\R,{\cal E})$ be the solution to the Cauchy problem
(\re{1dh}).
Then the asymptotics hold for $t\to\pm\infty$,
\be\la{qq}
\dot
q(t)=v_\pm+{\cal O}(|t|^{-1-\de}), ~~~~q(t)=v_\pm t+a_\pm+{\cal
O}(|t|^{-2\de}),
\ee
\be\la{S}
(E(x,t), A(x,t))=
(E_{v_\pm}(x-v_\pm t-a_\pm), A_{v_\pm}(x-v_\pm t-a_\pm))
+W^0(t)\bPsi_\pm+r_\pm(x,t) \ee with \be\la{rm} \Vert
r_\pm(t)\Vert_\cF=\cO(|t|^{-\de}). \ee
\end{theorem}

\noindent It suffices to  prove the asymptotics (\re{qq}), (\re{S})  for
$t\to+\infty$ since the system (\re{csh})-(\re{ph}) is time reversible.

%%%%%%%%%%%%%%%%%%%%%%%%%%%%%%%%%%%%%%%%%%%%%%%%%%%%%%%%%%%%%%%%%%%
%%%%%%%%%%%%%%%%%%%%%%%%%%%%%%%%%%%%%%%%%%%%%%%%%%%%%%%%%%%%%%%%%%%

\setcounter{equation}{0}

\section{Symplectic Projection}

%%%%%%%%%%%%%%%%%%%%%%%%%%%%%%%%%%%%%%%%%%%%%%%%%%%%%%%

\subsection{Symplectic Structure}

The system (\re{csh}) to (\re{ph}) reads as the Hamiltonian system \be\la{ham} \dot
Y=J{\cal D}{\cal H}(Y),\,\,\,J:=\left( \ba{cccc}
0 & E_3 & 0 & 0\\
-E_3 & 0 & 0 & 0\\
0 & 0 & 0 & E_3\\
0 & 0 & -E_3 & 0\\
\ea
\right),\,\,Y=(E,A,q,P)\in{\cal E},
\ee
where ${\cal D}{\cal H}$ is the Fr\'echet derivative of
 the Hamilton functional (\re{hamil}), $E_3$ is the $3\times3$ identity matrix.
Let us identify the tangent space to ${\cal E}$, at every point,
 with ${\cal E}$.
Consider the symplectic form $\Om$
 defined on
${\cal E}$ by \be\la{OmJ} \Om=\ds\int dE(x)\we dA(x)\,dx+dq\we dP,\,\,\,{\rm
i.e.}\,\,\,\Om(Y_1,Y_2)=\int(E_1\cdot A_2-E_2\cdot A_1)dx+q_1\cdot P_2-q_2\cdot P_1
\ee for $Y_k=(E_k,A_k,q_k,P_k)\in{\cal E}$, $k=1,2$ if the integral converges.

\begin{definition}
i) $Y_1\nmid Y_2$ means that $Y_1\in{\cal E}$ is symplectic orthogonal
to $Y_2\in{\cal E}$, i.e. $\Om(Y_1,Y_2)=0$.

ii) A projection operator $\bP:{\cal E}\to{\cal E}$
is called symplectic
 orthogonal if $Y_1\nmid Y_2$ for $Y_1\in\mbox{\rm Ker}\5\bP$ and
$Y_2\in\mbox \Im\bP$.
\end{definition}

%%%%%%%%%%%%%%%%%%%%%%%%%%%%%%%%%%%%%%%%%%%%%%%%%%%%%%%%%

\subsection{Symplectic Projection onto Solitary Manifold}

Let us consider the tangent space $\cT_{S(\si)}{\cal S}$ to
the manifold ${\cal S}$ at a point $S(\si)$.
The vectors $\tau_j:=\pa_{\si_j}S(\si)$, where $\pa_{\si_j}:=
\pa_{b_j}$ and $\pa_{\si_{j+3}}:=\pa_{v_{j}}$ with $j=1,2,3$,
form a basis in $\cT_{\si}{\cal S}$. In detail,
\be\la{inb}
\left.\ba{rclrrrrcrcl}
\tau_j=\tau_j(v)&:=&\pa_{b_j}S(\si)=
(&\!\!\!\!-\pa_j E_v(y)&\!\!\!\!,&\!\!\!\!-\pa_j A_v(y)&\!\!\!\!,
&\!\!e_j&\!\!\!\!,&\!\!0&\!\!\!\!)
\\
\tau_{j+3}=\tau_{j+3}(v)&:=&\pa_{v_j}S(\si)=(&\!\!\!\!\pa_{v_j}E_v(y)
&\!\!\!\!,&\!\!\!\!
\pa_{v_j}A_v(y)&\!\!\!\!,&\!\!0&\!\!\!\!,&\!\!
\pa_{v_j}P_v&\!\!\!\!) \ea\right|~~~j=1,2,3, \ee where  $y:=x-b$
is the {\it moving coordinate frame}, $e_1=(1,0,0)$ etc. Let us
stress that the functions $\tau_j$ will be considered always as
the functions of $y$, not of $x$.

By (\re{solw}) we have for the tangent vectors \be\la{solw1} \tau_j(v)\in{\cal
E}_{\alpha},\,\,\,{\rm for}\,\,\,\alpha<9/2,\,\,\,j=1,...,6. \ee

\begin{lemma}\la{Ome}
The matrix with the elements $\Om(\tau_l(v),\tau_j(v))$ is
non-degenerate for $|v|<1$.
\end{lemma}
The proof is made by a straightforward computation, see Appendix A.

Let us show that in a small neighborhood of the soliton manifold ${\cal S}$
a ``symplectic
orthogonal projection'' onto ${\cal S}$ is well-defined. Introduce the translations
$T_a:(\psi(\cdot),\pi(\cdot),q,p)\mapsto (\psi(\cdot-a),\pi(\cdot-a),q+a,p)$,
$a\in\R^3$. The manifold ${\cal S}$ is invariant with respect to the translations.
\begin{definition}
Put $v(Y):=P/\sqrt{1+P^2}$ where $P\in\R^3$ is the last component of the vector $Y$.
\end{definition}

\begin{lemma}\la{skewpro}
Let (\re{ro}) hold, $-9/2<\al$ and $\ov v<1$.
Then
\\
i) there exists a neighborhood ${\cal O}_\al({\cal S})$ of ${\cal
S}$ in ${\cal E}_\al$ and a map $\bPi:{\cal O}_\al({\cal
S})\to{\cal S}$  such that $\bPi$ is uniformly continuous on
${\cal O}_\al({\cal S})\cap\{Y\in\cE_\al:v(Y)\le\ov v\}$ in the
metric of ${\cal E}_\al$,
\be\la{proj}
\bPi Y=Y~~\mbox{for}~~
Y\in{\cal S}, ~~~~~\mbox{and}~~~~~ Y-S \nmid \cT_S{\cal
S},~~\mbox{where}~~S=\bPi Y.
\ee
ii) ${\cal O}_\al({\cal S})$ is
invariant with respect to the translations
 $T_a$, and
\be\la{commut}
\bPi T_aY=T_a\bPi Y,~~~~~\mbox{for}~~Y\in{\cal
O}_\al({\cal S}) ~~\mbox{and}~~a\in\R^3.
\ee
iii) For any $\ov
v<1$ there exists a $\ti v<1$ s.t.
 $|v(\bPi Y)|<\ti v$ when  $|v(Y)|<\ov v$.
\\\\
iv) For any $\ti v<1$ there exists an $r_\al(\ti v)>0$ s.t.
$S(\si)+Z\in\cO_\al(\cS)$ if $|v(S(\si))|<\ti v$ and $\Vert
Z\Vert_\al<r_\al(\ti v)$.
\end{lemma}
The proof is similar to that of Lemma 3.4 in \ci{IKV05}.

\medskip

We will call $\bPi$ the {\it symplectic orthogonal projection} onto ${\cal S}$.

\bc
The condition (\re{close})
implies that $Y_0=S+Z_0$ where $S=S(\si_0)=\bPi Y_0$, and
\be\la{closeZ}
\Vert Z_0\Vert_\beta \ll 1.
\ee
\ec

%%%%%%%%%%%%%%%%%%%%%%%%%%%%%%%%%%%%%%%%%%%%%%%%%%%%%%%%%%%%
%%%%%%%%%%%%%%%%%%%%%%%%%%%%%%%%%%%%%%%%%%%%%%%%%%%%%%%%%%%%

\setcounter{equation}{0}

\section{Linearization on the Solitary Manifold}

Let us consider a solution to the system (\re{csh})--(\re{ph}), and split it as
 the sum

\be\la{dec} Y(t)=S(\si(t))+Z(t), \ee where
$\si(t)=(b(t),v(t))\in\R^3\times\{|v|<1\}$ is an arbitrary smooth
function of
 $t\in\R$.
In detail, denote $Y=(E,A,q,P)$ and $Z=(e,a,r,\pi)$.
Then (\re{dec}) means that
\be \la{add}
\left.
\ba{rclrcl}
E(x,t)&=&E_{v(t)}(x-b(t))+e(x-b(t),t),
&q(t)&=&b(t)+r(t)\\
A(x,t)&=&A_{v(t)}(x-b(t))+a(x-b(t),t),
&P(t)&=&P_{v(t)}+\pi(t)
\ea
\right|
\ee
Let us
substitute (\re{add}) to (\re{csh})--(\re{ph}) and
linearize the equations in $Z$. Later we
 will choose $S(\si(t))=\bPi Y(t)$, i.e. $Z(t)$ is symplectic
orthogonal to $\cT_{S(\si(t))}{\cal S}$. However, this orthogonality condition is not
needed for the formal process of linearization. The orthogonality condition will be
important in Sections 6-7, where we derive ``modulation equations'' for the
parameters $\si(t)$.

Let us proceed to linearization. Setting $y=x-b(t)$ which is the {\it moving
coordinate frame}, we obtain from (\re{add}) and (\re{fsh})--(\re{ph}) that
\be\la{lin1E} \dot E = \dot v\cdot \na_v E_{v(t)}(y)-\dot b\cdot \na E_{v(t)}(y)+\dot
e(y,t)- \dot b\cdot \na e(y,t) = -\De(A_v(y)+a(y,t))-\Pi_s(\rho(y-r)\dot q), \ee
\be\la{lin1A} \dot A=\dot v\cdot \na_v A_{v(t)}(y)-\dot b\cdot \na A_{v(t)}(y)+\dot
a(y,t)-\dot b\cdot \na a(y,t)=-E_{v(t)}(y)-e(y,t), \ee \be\la{lin1q} \dot q=\dot
b+\dot
r=\fr{P_{v(t)}+\pi-\langle\rho(y-r),A_{v(t)}(y)+a(y,t)\rangle}{(1+(P_{v(t)}+\pi
-\langle\rho(y-r),A_{v(t)}(y)+a(y,t)\rangle)^2)^{1/2}},
\ee \be\la{lin1P} \dot P=\dot v\cdot \na_v
P_{v(t)}+\dot{\pi}=\langle\rho(y-r),\na(\dot q\cdot(A_{v(t)}(y)+a(y,t))\rangle. \ee
{\it Step i)} First we linearize the equation (\re{lin1q}). Note that
\be\la{ror}\rho(y-r)=\rho(y)-r\cdot\na\rho(y)+N_2(r),\ee where \be\la{N2} \Vert
N_2(r)\Vert_{0,\al}\le C_\al(\ov r)r^2, \ee uniformly in $|r|\le\ov r$ for any fixed
$\ov r$, for an arbitrary $\al>0$. Then (let us write $v$ instead of $v(t)$ and omit
the other arguments for simplicity) \be\la{N'2}
\langle\rho(y-r),A_v+a\rangle=\langle\rho,A_v\rangle+\langle\rho,a\rangle-\langle
r\cdot \na\rho,A_v\rangle+N'_2=\langle\rho,A_v\rangle+\langle\rho,a\rangle+N'_2, \ee
where $N'_2(r,a)=-\langle r\cdot \na\rho,a\rangle+\langle N_2,A_v+a\rangle$. Here we
use the equality $\langle r\cdot \na\rho,A_v\rangle=0$ which holds, since $A_v$ is
even and $\na\rho$ is odd. Further, since $P_v-\langle\rho,A_v\rangle=p_v$ by
(\re{sosol}), we get
$P_v+\pi-\langle\rho(y-r),A_v+a\rangle=P_v+\pi-\langle\rho,A_v\rangle-
\langle\rho,a\rangle-N'_2=p_v+\pi-\langle\rho,a\rangle-N'_2=p_v+s$,
where $s:=\pi-\langle\rho,a\rangle-N'_2$. Applying Taylor expansion we obtain
$$
(1+(p_v+s)^2)^{-1/2}=\fr{1}{(1+p_v^2)^{1/2}}-\fr{p_v\cdot s}{(1+p_v^2)^{3/2}}+N_3'
=\fr{1}{(1+p_v^2)^{1/2}}-\fr{v\cdot s}{1+p_v^2}+N_3',
$$
since $p_v/(1+p_v^2)^{1/2}=v$. Finally,
$$
\fr{P_{v}+\pi-\langle\rho(y-r),A_{v}+a\rangle}{(1+(P_{v}+\pi-\langle\rho(y-r),A_{v}(y)
+a\rangle)^2)^{1/2}}=\fr{p_v+s}{(1+(p_{v}+s)^2)^{1/2}}
$$
$$
=v-\fr{(v,s)v}{(1+(p_{v})^2)^{1/2}}+\fr{s}{(1+(p_{v})^2)^{1/2}}+N_3''=
v+\nu(s-(v\cdot s)v)+N_3'',
$$
where $\nu=(1-v^2)^{1/2}=(1+p_v^2)^{-1/2}$. Insert the expression for $s$, then
the equation (\re{lin1q}) becomes \be\la{lin-r} \dot r=v-\dot
b+B_v(\pi-\langle\rho,a\rangle)+N_3, \ee where $B_v:=\nu(E-v\otimes v)$, and
\be\la{N3} |N_3(Z)|\le C(\ti v)\Vert Z\Vert^2_{-\al} \ee
 uniformly in $|v|\le\ti v<1$, for an arbitrary $\al>0$.

\medskip

\noindent{\it Step ii)} Next we linearize the equation (\re{lin1E}). By (\re{ror})
and (\re{lin-r}) we obtain
$$
\rho(y-r)\dot q =\rho v+\rho
B_v(\pi-\langle\rho,a\rangle)-r\cdot\na\rho v+N_1'.
$$
Substitute to the equation
(\re{lin1E}) and take (\re{solv}) into account, then we get \be\la{lin-e} \dot e=\dot
b\cdot\na e-\De a+(\dot b-v)\cdot\na E_v-\dot v\cdot\na_v E_v -\Pi_s(\rho
B_v(\pi-\langle\rho,a\rangle)-r\cdot\na\rho v)+N_1, \ee where for $N_1$ the same
bound holds,
\be\la{N1bound}
\Vert N_1(Z)\Vert\le C(\ti v)\Vert Z\Vert_{-\al}^2,\,\,\,\forall\,\al>0.
\ee

\medskip

\noindent
{\it Step iii)} Further, by (\re{solv}) the equation (\re{lin1A}) becomes
\be\la{lin-a} \dot a=-e+\dot b\cdot\na a+(\dot b-v)\cdot\na A_v-\dot v\cdot\na_v A_v.
\ee

\medskip

\noindent
{\it Step iv)} Let us proceed to the equation (\re{lin1P}). We have
$$
\dot P=\dot v\cdot\na_v P_v+\dot\pi=\langle\rho(y-r),\na(\dot q\cdot(A_v+a))\rangle
$$
$$
=\langle\rho-r\cdot\na\rho+N_2,\na((v+B_v(\pi-\langle\rho,a\rangle)+
N_3)\cdot(A_v+a))\rangle =\langle\rho,v\cdot\na a\rangle-\langle
r\cdot\na\rho,\na(v\cdot A_v)\rangle+N_4,
$$
since $\langle\rho,v\cdot\na A_v\rangle=0$ and
$\langle\rho,B_v(\pi-\langle\rho,a\rangle)\cdot\na A_v\rangle=0$. Finally, the
equation becomes
\be\la{lin-p}
\dot\pi=\langle\rho,\na(v\cdot a)\rangle-\langle r\cdot\na\rho,\na(v\cdot A_v)\rangle-\dot
v\cdot\na_v P_v+N_4,
\ee
where for
$N_4(v,Z)$ the estimate like (\re{N3}) holds. We write the equations
(\re{lin-r}), (\re{lin-e})--(\re{lin-p}) as \be\la{lin} \dot
Z(t)=A(t)Z(t)+T(t)+N(t),\,\,\,t\in\R. \ee
%%%%%%%%%%%%%%%%%%%%%%%%%%%%%%%%
%%%%%%%%%%%%%%%%%%%%%%%%%%%%%%%%%%%%%%%%%%%
Here the operator $A(t)$ depends on $\si(t)=(b(t),v(t))$. We will use the parameters
$v=v(t)$ and $w:=\dot b(t)$. Then $A(t)=A_{v,w}$ can be written in the form
\be\la{AA}
A_{v,w} \left( \ba{c} e \\ a \\ r \\ \pi \ea \right):= \left( \ba{cccc}
w \cdot\na & -\De+\Pi_s(\rho B_v\langle\rho,\cdot\rangle) & \Pi_s(\cdot\na\rho v) &
-\Pi_s(\rho B_v\cdot) \\
-1 & w\cdot \na & 0 & 0 \\
0 & -B_v\langle\rho,\cdot\rangle & 0 & B_v \\
0 & \langle\rho,\na(v\cdot)\rangle & -\langle\cdot\na\rho,\na(v\cdot A_v)\rangle & 0
\ea
\right)\left(
\ba{c}
e \\ a \\ r \\ \pi
\ea
\right).
\ee
%%%%%%%%%%%%%%%%%%%%%%%%%%%%%%%%%%%%%%%%%%%%%%%%%%%%%%%%%%%
%%%%%%%%%%%%%%%%%%%%%%%%%%%%%%%%%%%%%%%%%%%%%%%%%%%%%%%%%%%
Furthermore,   $T(t)$ and $N(t)$ in  (\re{lin}) stand for
\be\la{TN}
T(t)=T_{v,w}=\left(
\ba{c}
(w-v)\cdot\na E_v-\dot v\cdot\na_v E_v\\
(w-v)\cdot\na A_v-\dot v\cdot\na_v A_v\\
v-w \\
-\dot v\cdot\na_v P_v \ea \right),\,\,\,\,N(t)=N(v,Z)=\left( \ba{l} N_1(v,Z) \\
0 \\ N_3(v,Z) \\ N_4 (v,Z) \ea \right), \ee where $v=v(t)$, $w=w(t)$,
 and $Z=Z(t)$. The estimates (\re{N2}), (\re{N3}), and (\re{N1bound}) imply the
 following
\bl\la{Nbound}
For any $\alpha>0$
\be\la{N14} \Vert N(v,Z)\Vert_\al\le
C(\ti v)\Vert Z\Vert_{-\al}^2 \ee uniformly in $v$ and $Z$ with $\Vert
Z\Vert_{-\al}\le r_{-\al}(\ti v)$ and $|v|<\ti v<1$.
\el
\brs\la{rT} {\rm i) The term $A(t)Z(t)$ in the right hand side of the equation
(\re{lin}) is linear  in $Z(t)$, and $N(t)$ is a {\it high order term} in $Z(t)$.
\\
ii)
Formulas (\re{inb}) and (\re{TN}) imply:
\be\la{Ttang}
T(t)=-\sum\limits_{l=1}^3[(w-v)_l\tau_l+\dot v_l\tau_{l+3}]
\ee
and hence $T(t)\in \cT_{S(\si(t))}{\cal S}$, $t\in\R$.
The
term $T(t)$ vanishes if $S(\si(t))$ is a soliton solution since in this case $\dot
v=0$ and $w=\dot b=v$. Otherwise $T(t)$ is a {\it zero order term} which does not vanish
although $S(\si(t))$ belongs to the solitary manifold.
In our context we will show that
$T(t)$ rapidly decays as $t\to\infty$ (see (\re{Tta}) below).
}
\ers

%%%%%%%%%%%%%%%%%%%%%%%%%%%%%%%%%%%%%%%%%%%%%%%%%%%%%%%%%%%%%%
%%%%%%%%%%%%%%%%%%%%%%%%%%%%%%%%%%%%%%%%%%%%%%%%%%%%%%%%%%%%%%

\setcounter{equation}{0}
\section{The Linearized Equation}

Here we study some properties of the
operator (\re{AA}). First, let us
compute the action of $A_{v,w}$ on the tangent vectors $\tau_j$ to the solitary
manifold ${\cal S}$.

\begin{lemma} \la{ljf}
The operator $A_{v,w}$ acts on the
tangent vectors $\tau_j(v)$ to the solitary manifold
as follows,
\be\la{Atanform}
A_{v,w}[\tau_j(v)]=(w-v)\cdot\na\tau_j(v),\,\,\,A_{v,w}[\tau_{j+3}(v)]=
(w-v)\cdot\na\tau_{j+3}(v)+\tau_j(v),\,\,\,j=1,2,3.
\ee
\end{lemma}
{\bf Proof } To get (\re{Atanform}), differentiate the stationary equations
(\re{solv}) in $x_j$ and $v_j$, cf. \ci{IKV05}. \hfill$\Box$

\smallskip

\noindent Consider the linear equation \be\la{line} \dot
X(t)=A_{v,w}X(t),~~~~~~~t\in\R \ee with an arbitrary fixed $v$ such that $|v|<1$ and
$w\in\R^3$. Let us define the space
$$
{\cal E}^+=H^1_s \oplus \dot H^2_s\oplus \R^3 \oplus \R^3
$$

\begin{lemma} \la{haml}
i) For any $v$, $|v|<1$, $w\in\R^3$ the equation (\re{line})
formally
 can be written as the Hamiltonian system  (cf. (\re{ham})),
\be\la{lineh}
\dot X(t)=
JD{\cal H}_{v,w}(X(t)),~~~~~~~t\in\R,
\ee
where $D{\cal H}_{v}$ is the Fr\'echet derivative of the
Hamilton functional
$$
{\cal H}_{v,w}(X)=\fr12\int\Big[|e|^2+|\na a|^2\Big]dy+\int a(w\cdot\na)e dy+
\fr12(B_v\langle\rho,a\rangle)\cdot\langle\rho,a\rangle
$$
\be\la{H0}
+\fr12\pi\cdot B_v\pi+\langle r\cdot\na\rho v,a\rangle-\langle\rho B_v\pi,a\rangle+
\fr12\langle r\cdot\na\rho,v\cdot(\na A_v)\rangle,~~~~X=(e,a,r,\pi)\in \cE.
\ee
\\
ii) Energy conservation law holds
for the solutions $X(t)\in C(\R,\cE)$,
\be\la{enec}
\cH_{v,w}(X(t))=\co,~~~~~t\in\R.
\ee
iii) The skew-symmetry relation holds,
\be\la{com}
\Omega(A_{v,w}X_1,X_2)=-\Omega(X_1,A_{v,w}X_2), ~~~~X_1\in\cE,\,\,X_2\in\cE^+.
\ee
\end{lemma}
The proof is similar to that in \ci{IKV05}. We will apply Lemma \re{haml} mainly to
the operator $A_{v,v}$ corresponding to $w=v$. In that case the linearized equation
has the following additional essential features.

\bl\la{ceig} Let us assume that $w=v$ and $|v|<1$. Then
\\
i) The tangent vectors $\tau_j(v)$ with $j=1,2,3$ are eigenvectors, and
$\tau_{j+3}(v)$ are root vectors of the operator $A_{v,v}$, corresponding to zero
eigenvalue, i.e. \be\la{Atanformv} A_{v,v}[\tau_j(v)]=0,\,\,\,A_{v,v}[\tau_{j+3}(v)]=
\tau_j(v),\,\,\,j=1,2,3. \ee ii) The Hamilton function (\re{H0}) is positive
definite, \be\la{H0vv} {\cal H}_{v,v}(X)\ge 0. \ee \el
{\bf Proof } The first
statement follows from (\re{Atanform}). To prove the second statement note that for
$X=(e,a,r,\pi)\in \cE$ one has
$$
{\cal H}_{v,v}(X)=\fr12\int\Big[|e|^2+|\na a|^2\Big]dy+\int a(v\cdot\na)e dy+
\fr12(B_v\langle\rho,a\rangle)\cdot\langle\rho,a\rangle
$$
$$
+\fr12\pi\cdot B_v\pi+\langle r\cdot\na\rho v,a\rangle-\langle\rho B_v\pi,a\rangle+
\fr12\langle r\cdot\na\rho,v\cdot(\na A_v)\rangle
$$
$$
=\fr12 (B_v(\pi-\langle\rho,a\rangle))\cdot(\pi-\langle\rho,a\rangle)+
\fr12(\langle e,e\rangle+\langle(v\cdot\na)a,(v\cdot\na)a\rangle-
\langle e,(v\cdot\na)a\rangle)
$$
$$
+\fr12(\langle(-\De+(v\cdot\na)^2)a,a\rangle+\langle(r\cdot\na)\rho v,a\rangle+
\langle(r\cdot\na)\rho,v\cdot(\na A_v)\rangle).
$$
Here the first line is clearly nonnegative, since $B_v$ is nonnegative definite.
The last line in Fourier space by (\re{Sfur}) equals
$$
\fr12\int\left((k^2-(kv)^2)|\hat a|^2-2i(kr)\hat\rho(v\cdot\ov{\hat a})+
\fr{(kr)^2|\hat\rho|^2v^2}{k^2-(kv)^2}\right)dk.
$$
The integrand is nonnegative, since $|\Re[i(kr)\hat\rho(v\cdot\ov{\hat
a})]|\le|(kr)||\hat\rho||v||\hat a|$.\hfill$\Box$
\br {\rm
 For a soliton solution of the system
(\re{csh})--(\re{ph})
 we have $\dot b=v$, $\dot v=0$, and hence $T(t)\equiv 0$.
Thus, the equation
(\re{line})
is the linearization of the system
(\re{csh})--(\re{ph}) on a soliton solution. In fact, we do not linearize
 (\re{csh})--(\re{ph}) on a soliton solution, but on a trajectory
$S(\si(t))$ with $\si(t)$
 being nonlinear in $t$. We will show later that $T(t)$ is quadratic in $Z(t)$
if we choose
$S(\si(t))$ to be  the symplectic orthogonal projection of $Y(t)$.
Then
(\re{line})  is again the linearization of (\re{csh})--(\re{ph}).
}
\er

%%%%%%%%%%%%%%%%%%%%%%%%%%%%%%%%%%%%%%%%%%%%%%%%%%%%%%%%%%%%%%%%%%
%%%%%%%%%%%%%%%%%%%%%%%%%%%%%%%%%%%%%%%%%%%%%%%%%%%%%%%%%%%%%%%%%%

\setcounter{equation}{0}

\section{Symplectic Decomposition of the Dynamics}

Here we decompose the dynamics in two components: along the manifold
 ${\cal S}$ and in transversal directions. The equation (\re{lin})
is obtained without any assumption on $\si(t)$ in (\re{dec}).
We are going to choose $S(\si(t)):=\bPi Y(t)$, but then we need
to know  that
\be\la{YtO}
Y(t)\in \cO_\al(\cS),~~~~~t\in\R,
\ee
with some $\cO_\al(\cS)$ defined in Lemma \re{proj}.
It is true for $t=0$ and $\al=\beta$ by our main assumption
 (\re{close}) with sufficiently small $d_\beta>0$.
Then  $S(\si(0))=\bPi Y(0)$ and  $Z(0)=Y(0)-S(\si(0))$
are well defined.
We will prove below that (\re{YtO}) holds
with $\al=-\beta$
 if $d_\beta$ is sufficiently small.
First, the a priori estimate (\re{ov-v})
together with Lemma \re{skewpro} iii)
imply that $\bPi Y(t)=S(\si(t))$ with $\si(t)=(b(t),v(t))$,
 and
\be\la{vsigmat}
|v(t)|\le \ti v<1,~~~~~~t\in\R
\ee
if $Y(t)\in \cO_{-\beta}(\cS)$. Denote by $r_{-\beta}(\ti v)$ the positive
number from
Lemma \re{skewpro} iv) which corresponds to $\al=-\beta$.
Then $S(\si)+Z\in \cO_{-\beta}(\cS)$ if
$\si=(b,v)$ with $|v|<\ti v$ and
$ \Vert Z\Vert_{-\beta}<r_{-\beta}(\ti v)$.
Note that (\re{ov-v}) implies
$\Vert Z(0)\Vert_{-\beta}<r_{-\beta}(\ti v)$ if
 $d_\beta$ is sufficiently small.
Therefore, $S(\si(t))=\bPi Y(t)$ and  $Z(t)=Y(t)-S(\si(t))$
are well defined for
$t\ge 0$
so small that
 $\Vert Z(t)\Vert_{-\beta} < r_{-\beta}(\ti v)$.
This is formalized by the following standard definition.
\begin{definition}
$t_*$ is the ``exit time'', \be\la{t*} t_*=\sup \{t>0: \Vert
Z(s)\Vert_{-\beta} < r_{-\beta}(\ti v),~~0\le s\le
t\},\,\,\,Z(s)=Y(s)-S(\si(s)). \ee
\end{definition}
One of our main goals is to prove that $t_*=\infty$ if $d_\beta$ is sufficiently
small. This would follow if we show that \be\la{Zt} \Vert
Z(t)\Vert_{-\beta}<r_{-\beta}(\ti v)/2,~~~~~0\le t < t_*. \ee Note that \be\la{Qind}
|r(t)|\le\ov r:= r_{-\beta}(\ti v), ~~~~~0\le t< t_*. \ee Now $N(t)$ in (\re{lin})
satisfies, by (\re{N14}) with $\al=-\beta$, the following estimate, \be\la{Nest} \Vert
N(t)\Vert_{\beta}\le C_\beta(\ti v)\Vert Z(t)\Vert^2_{-\beta}, \,\,\,0\le t<t_*. \ee

%%%%%%%%%%%%%%%%%%%%%%%%%%%%%%%%%%%%%%%%%%%%%%%%%%%%%%%%%%%%%%%%%

\setcounter{equation}{0}

\section{Longitudinal Dynamics: Modulation Equations}

>From now on
 we fix the decomposition
$Y(t)=S(\si(t))+Z(t)$ for $0<t<t_*$
by setting $S(\si(t))=\bPi Y(t)$ which is equivalent to the
 symplectic orthogonality condition of type (\re{proj}),
\be\la{ortZ}
Z(t)\nmid\cT_{S(\si(t))}{\cal S},\,\,\,0\le t<t_*.
\ee
This allows us to simplify drastically the
asymptotic analysis of the
dynamical equations (\re{lin})
 for the transversal component $Z(t)$. As the first step, we derive the
longitudinal dynamics, i.e. the ``modulation equations'' for the parameters
 $\si(t)$.
Let us derive a system of ordinary differential equations for the
vector $\si(t)$. For this purpose,
let us write (\re{ortZ}) in the form
\be\la{orth}
\Om(Z(t),\tau_j(t))=0,\,\,j=1,\dots,6, ~~~~~~~0\le t<t_*,
\ee
where the vectors $\tau_j(t)=\tau_j(\si(t))$ span the
tangent space
$\cT_{S(\si(t))}{\cal S}$.
Note that $\si(t)=(b(t),v(t))$, where
\be\la{sit}
|v(t)|\le \ti v<1,~~~~~~~~~0\le t<t_*,
\ee
by Lemma \re{skewpro} iii).
  It would be convenient for us to use some other
parameters $(c,v)$ instead of $\si=(b,v)$, where
$c(t)=
b(t)-\ds\int^t_0 v(\tau)d\tau$ and
\be\la{vw}
\dot c(t)=\dot b(t)-v(t)=w(t)-v(t), ~~~~~~~~~0\le t<t_*.
\ee
We do not
need an explicit form of the equations for $(c,v)$ but the following
 statement.

\begin{lemma}\la{mod}({\rm cf. \ci{IKV05}, Lemma 6.2})
Let $Y(t)$ be a solution to the Cauchy problem (\re{1dh}), and (\re{dec}),
(\re{orth}) hold. Then $(c(t),v(t))$ satisfies the equation \be\la{parameq} \left(
\ba{l} \dot c(t) \\ \dot v(t) \ea \right)={\cal N}(\si(t),Z(t)), ~~~~~~~0\le t<t_*,
\ee where \be\la{NZ} {\cal N}(\si,Z)={\cal O}(\Vert Z\Vert^2_{-\beta}) \ee uniformly
in  $\si\in\{(b,v):|v|\le\ti v\}$.
\end{lemma}
{\bf Proof } We differentiate (\re{orth}) in $t$ and take the equation (\re{lin})
into account. Then (see details of computation in \ci{IKV05}, Lemma 6.2) we obtain,
in the vector form \ci[(6.18)]{IKV05}: \be\la{modular} 0=\Om(v)\left( \ba{c} \dot c
\\ \dot v \ea \right)+{\cal M}_0(\si,Z)\left( \ba{c} \dot c \\ \dot v \ea
\right)+{\cal N}_0(\si,Z),\,\,\,{\cal N}_{0j}(\si,Z)=\Om(N,\tau_j). \ee Here the
matrix $\Om(v)$ has the matrix elements $\Om(\tau_l,\tau_j)$ and hence is invertible
by Lemma \re{Ome}. The $6\times6$ matrix ${\cal M}_0(\si,Z)$ has the matrix elements
$\sim\,\,\Vert Z\Vert_{-\beta}$ and hence we can resolve the equation (\re{modular})
with respect to $(\dot c,\dot v)$. Then (\re{NZ}) follows from Lemma \re{Nbound} with
$\al=\beta$, since ${\cal N}_0={\cal O}(\Vert Z\Vert_{-\beta}^2)$. \hfill $\Box$

\br \la{radiab} {\rm The equations
(\re{parameq}), (\re{NZ}) imply that the soliton parameters $c(t)$ and $v(t)$ are
{\it adiabatic invariants} (see \ci{AKN}). } \er

%%%%%%%%%%%%%%%%%%%%%%%%%%%%%%%%%%%%%%%%%%%%%%%%%%%%%%%%%%%
%%%%%%%%%%%%%%%%%%%%%%%%%%%%%%%%%%%%%%%%%%%%%%%%%%%%%%%%%%%

\setcounter{equation}{0}

\section{Decay for the Transversal Dynamics}
Here we prove
 the
following time decay of the transversal component $Z(t)$:
\bp\la{pdec} Let all
conditions of Theorem \re{main} hold. Then $t_*=\infty$, and
\be\la{Zdec}
\Vert
Z(t)\Vert_{-\beta}\le \ds\fr {C(\rho,\ti
v,d_\beta)}{(1+|t|)^{1+\de}},~~~~~t\ge0.
\ee
\ep
In next Section  we will show that our main Theorem \re{main} can be
derived from the transversal decay (\re{Zdec}).
We will derive this decay
from our equation (\re{lin}) for the
transversal component $Z(t)$. This equation can be specified using Lemma \re{mod}.
Namely,
by (\re{Ttang}) and (\re{vw})
$$
T(t)=-\sum\limits_{l=1}^3[\dot c_l\tau_l+\dot v_l\tau_{l+3}].
$$
Then Lemma \re{mod} implies that
\be\la{Tta} \Vert T(t)\Vert_{\beta}\le C(\ti v)\Vert Z(t)\Vert^2_{-\beta},
~~~~~~~~~0\le t<t_*. \ee
Note that the norm $\Vert T(t)\Vert_{\beta}$ is well defined by the condition
(\re{neutr}). Thus, in (\re{lin}) we should combine the terms $T(t)$ and $N(t)$ and
obtain
\be\la{reduced} \dot Z(t)=A(t)Z(t)+\ti N(t), ~~~~~~~~~0\le t<t_*, \ee
where $A(t)=A_{v(t),w(t)}$, and $\ti N(t):=T(t)+N(t)$. By (\re{Tta}) and (\re{Nest})
we have
\be\la{redN} \Vert\ti N(t)\Vert_{\beta}\le C\Vert Z(t)\Vert^2_{-\beta},~~~~
~~~~~~~~~0\le t<t_*. \ee In all remaining part of our paper we will analyze mainly
the {\bf basic equation} (\re{reduced}) to establish the decay (\re{Zdec}). We are
going to derive the decay using the bound (\re{redN}) and
 the
orthogonality condition  (\re{ortZ}).

Let us comment on two main difficulties in proving (\re{Zdec}). The difficulties are
common for the problems studied in \ci{BP2,Cu}. First, the linear part of the
equation is non-autonomous, hence we cannot apply directly known methods of
scattering theory. Similarly to the approach of  \ci{BP2,Cu}, we reduce the problem
to the analysis of the {\it frozen} linear equation, \be\la{Avv} \dot X(t)=A_1X(t),
~~t\in\R, \ee where $A_1$ is the operator $A_{v_1,v_1}$ defined in (\re{AA}) with
$v_1=v(t_1)$ and a fixed $t_1\in[0,t_*)$. Then we estimate the error by the method of
majorants.

Second, even for the frozen equation (\re{Avv}), the decay of type  (\re{Zdec}) for
all solutions does not hold without  the orthogonality condition  of type
(\re{ortZ}). Namely, by (\re{Atanformv}) the equation (\re{Avv}) admits the {\it
secular solutions} \be\la{secs} X(t)=\sum_1^3 C_{j}\tau_j(v_1)+\sum_1^3
D_j[\tau_j(v_1)t+\tau_{j+3}(v_1)] \ee which arise also by differentiation of the
soliton (\re{sosol}) in the parameters $a$ and $v_1$ in the moving coordinate
$y=x-v_1t$. Hence, we have to take into account the orthogonality condition
(\re{ortZ}) in order to avoid the secular solutions. For this purpose we will apply
the corresponding symplectic orthogonal projection which kills the ``runaway
solutions''  (\re{secs}).

\bd i) Denote by $\bPi_v$, $|v|<1$, the symplectic orthogonal
projection of ${\cal E}$ onto the tangent space $\cT_{S(\si)}{\cal
S}$, and
 $\bP_v=\bI-\bPi_v$.
\\
ii) Denote by $\cZ_v=\bP_v\cE$ the space symplectic
orthogonal to $\cT_{S(\si)}{\cal S}$ with
$\si=(b,v)$ (for an arbitrary $b\in\R$).
\ed
Note that by the linearity,
\be\la{Piv}
\bPi_vZ=\sum\bPi_{jl}(v)
\tau_j(v)\Om(\tau_l(v),Z),~~~~~~~~~~Z\in\cE,
\ee
 with some smooth coefficients $\bPi_{jl}(v)$.
Hence, the projector $\bPi_v$, in the variable $y=x-b$,
does not depend on $b$, and this explains the choice
of the subindex in $\bPi_v$ and $\bP_v$.

Now we have the symplectic orthogonal decomposition
\be\la{sod}
\cE_\beta=\cT_{S(\si)}{\cal S}+\cZ_v,~~~~~~~\si=(b,v),
\ee
and the symplectic orthogonality  (\re{ortZ})
can be written in the following equivalent forms,
\be\la{PZ}
\bPi_{v(t)} Z(t)=0,~~~~\bP_{v(t)}Z(t)= Z(t),~~~~~~~~~0\le t<t_*.
\ee

\br\la{rZ}
{\rm
The tangent space $\cT_{S(\si)}{\cal S}$ is invariant under
the operator $A_{v,v}$ by Lemma \re{ceig} i), hence
the space  $\cZ_v$ is also invariant by
(\re{com}): $A_{v,v}Z\in \cZ_v$
for {\it sufficiently smooth}  $Z\in \cZ_v$.
}
\er

The following
 proposition  is one of main ingredients for proving
(\re{Zdec}). Let us consider the Cauchy problem for the  equation
(\re{Avv}) with $A_1=A_{v_1,v_1}$ for a fixed $v_1$, $|v_1|<1$. Recall that
$\beta=4+\de$, $0<\de<1/2$.
\begin{pro}\la{lindecay}
Let the Wiener condition (\re{W}) and the condition
(\re{neutr}) hold, $|v_1|\le\ti v<1$, and $X_0\in\cE$. Then
\\
i) The equation (\re{Avv}), with $A_1=A_{v_1,v_1}$, admits the unique solution
 $e^{A_1t}X_0:=X(t)\in C_b(\R, \cE)$
with the initial condition $X(0)=X_0$.
\\
ii) For $X_0\in \cZ_{v_1}\cap \cE_\beta$, the
the following decay holds,
\be\la{frozenest}
\Vert
e^{A_1t}X_0\Vert_{-2-\de}\le \fr{C(\ti v)}{(1+|t|)^{1+\de}}
\Vert X_0\Vert_{\beta},~~~~~t\in\R.
\ee
\end{pro}

Part i) follows by standard arguments using the positivity (\re{H0vv}) of the
Hamilton functional.
Part ii) will be proved in Sections 10-13
developing general strategy \ci{IKV05}. Namely,
the equation (\re{Avv}) is a system of four
equations involving field components, $E$ and $A$ as well as
vector components, $r$ and $\pi$. We apply Fourier-Laplace
transform, express the field components in terms of the vector
components from the first two equations and substitute to the
third and the fourth equations. Then we obtain a closed system for
the vector components alone and prove their decay. Finally, for
the field components we come to a wave equation with a right hand
side which has the established decay. This implies the
corresponding decay for the field components.

%%%%%%%%%%%%%%%%%%%%%%%%%%%%%%%%%%%%%%%%%%%%%%%%%%%%%%%%%%%

\subsection{Frozen Form of Transversal Dynamics}

Now  let us fix an arbitrary $t_1\in [0,t_*)$, and rewrite the equation
(\re{reduced}) in a ``frozen form'' \be\la{froz} \dot Z(t)=A_1Z(t)+(A(t)-A_1)Z(t)+\ti
N(t),\,\,\,~~~~0\le t<t_*, \ee where $A_1=A_{v(t_1),v(t_1)}$ and \be\la{frozendif}
\!\!\!\!\!\!\!\!\!\!\! A(t)\!-\!A_1\!=\!\!\left( \ba{cccc} [w\!-\!v_1]\cdot \na &
\Pi_s(\rho(B_{v}-B_{v_1})\langle\rho,\cdot\rangle) & \Pi_s(\cdot\na\rho(v-v_1)) &
-\Pi_s(\rho(B_{v}-B_{v_1})\cdot) \\
0 & [w\!-\!v_1]\cdot \na & 0 & 0 \\
0 & -(B_{v}-B_{v_1})\langle\rho,\cdot\rangle & 0 & B_{v}\!-\!B_{v_1} \\
0 & \langle\rho,(v-v_1)\na\cdot\rangle & -\langle\cdot\na\rho,(v\na A_{v}-
v_1\na A_{v_1})\rangle & 0
\ea \right)
\ee
where $w=w(t)$, $v=v(t)$, $v_1=v(t_1)$.
The next trick is important since it allows us to kill the ``bad terms''
 $[w(t)\!-\!v(t_1)]\cdot \na$ in the operator $A(t)-A_1$.
\begin{definition}\la{d71}
Let us change the  variables $(y,t)\mapsto (y_1,t)=(y+d_1(t),t)$
where
\be\la{dd1}
d_1(t):=\int_{t_1}^t(w(s)-v(t_1))ds, ~~~~0\le t\le t_1.
\ee
\end{definition}
Next define
\beqn\la{Z1}
Z_1(t)&=&(e_1(y_1,t),a_1(y_1,t),r(t),\pi(t)):=
(e(y,t),a(y,t),r(t),\pi(t))\nonumber\\
&=&
(e(y_1-d_1(t),t),a(y_1-d_1(t),t),r(t),\pi(t)).
\eeqn
Then we obtain the final form of the
``frozen equation'' for the transversal dynamics
\be\la{redy1}
\dot Z_1(t)=A_1Z_1(t)+B_1(t)Z_1(t)+N_1(t),\,\,\,0\le t\le t_1,
\ee
where $N_1(t)=\ti N(t)$ is expressed in terms of $y=y_1-d_1(t)$, and
$$
B_1(t)=\left(
\ba{cccc}
0 & \Pi_s(\rho(B_{v(t)}-B_{v(t_1)})\langle\rho,\cdot\rangle) &
\Pi_s(\cdot\na\rho(v(t)-v(t_1))) & -\Pi_s(\rho(B_{v(t)}-B_{v(t_1)})\cdot) \\
0 & 0 & 0 & 0 \\
0 & -(B_{v(t)}-B_{v(t_1)})\langle\rho,\cdot\rangle & 0 & B_{v(t)}\!-\!B_{v(t_1)} \\
0 & \langle\rho,(v(t)-v(t_1))\cdot\na\cdot\rangle &
-\langle\cdot\na\rho,(v(t)\na A_{v(t)}-v(t_1)\na A_{v(t_1)})\rangle & 0
\ea \right)
$$
Let us derive appropriate bounds for the
``remainder terms'' $B_1(t)Z_1(t)$ and $N_1(t)$ in (\re{redy1}).

\begin{lemma}\la{cor1}{\rm \ci[Corollary 7.3 and 7.4]{IKV05}}
The following  bounds hold \be\la{N1est} \Vert N_1(t)\Vert_{\beta}\le\Vert
Z_1(t)\Vert^2_{-\beta} (1+|d_1(t)|)^{3\beta}~,~~~~~~0\le t\le t_1, \ee

\be\la{B1Z1est} \Vert B_1(t)Z_1(t)\Vert_{\beta}\le C \Vert
Z_1(t)\Vert_{-\beta}\int_t^{t_1}(1+|d_1(\tau)|)^{2\beta} \Vert
Z_1(\tau)\Vert^2_{-\beta} d\tau~,~~~~~~0\le t\le t_1. \ee
\end{lemma}

%%%%%%%%%%%%%%%%%%%%%%%%%%%%%%%%%%%%%%%%%%%%%%%%%%%%%%%%%%%%%%%%%%%%%%%%%%%%%

\subsection{Integral Inequality}

The equation (\re{redy1}) can be written in the integral form:
\be\la{Z1duh}
Z_1(t)=e^{A_1t}Z_1(0)+\int_0^te^{A_1(t-s)}[B_1Z_1(s)+N_1(s)]ds,\,\,\,
0\le t\le t_1.
\ee
We apply the symplectic orthogonal
projection $\bP_1:=\bP_{v(t_1)}$ to both sides, and get
$$
\bP_1Z_1(t)=e^{A_1t}\bP_1Z_1(0)+\int_0^te^{A_1(t-s)}\bP_1[B_1Z_1(s)+N_1(s)]ds.
$$
We have used here that $\bP_1$ commutes
with
the group $e^{A_1t}$ since the space $\cZ_1:=\bP_1\cE$ is invariant
with respect to $e^{A_1t}$, see Remark \re{rZ}.
Applying (\re{frozenest})
 we obtain that
\be\la{bPZ} \Vert
\bP_1Z_1(t)\Vert_{-2-\de}\le\fr{C}{(1+t)^{1+\de}} \Vert
\bP_1Z_1(0)\Vert_{\beta}+C\int_0^t\fr1{(1+|t-s|)^{1+\de}}\Vert
 \bP_1[B_1Z_1(s)+N_1(s)]\Vert_{\beta}ds.
\ee The operator $\bP_1=\bI-\bPi_1$ is continuous in $\cE_\beta$
by (\re{Piv}). Hence,  from (\re{bPZ}) and (\re{N1est}),
(\re{B1Z1est}), we obtain that \beqn\la{duhest}
\!\!\!\!\!\!\!\!\!\!\!\!\!\!\!\!\!\!&&\!\!\!\!\!\! \Vert
\bP_1Z_1(t)\Vert_{-2-\de}\le\fr{C}{(1+t)^{1+\de}}\Vert
Z_1(0)\Vert_{\beta}
\nonumber\\
\!\!\!\!\!\!\!\!\!\!\!\!\!\!\!\!\!\!&&\!\!\!\!\!\! +C(\ov
d_1)\int_0^t\fr1{(1+|t-s|)^{1+\de}}\left[\Vert Z_1(s)\Vert_{-\beta}
\int_s^{t_1}\Vert Z_1(\tau)\Vert^2_{-\beta}d\tau+ \Vert
Z_1(s)\Vert^2_{-\beta}\right]ds,\,\,\,0\le t\le t_1, \eeqn where $\ov
d_1:=\sup_{0\le t\le t_1} |d_1(t)| $. Since $\Vert
Z_1(t)\Vert_{\pm\beta}\le C(\ov d_1) \Vert Z(t)\Vert_{\pm\beta}$, we can rewrite
(\re{duhest}) as \beqn\la{duhestr} \!\!\!\!\!\!\!\!\!\!\!\!\!\!\!\!\!\!&&\!\!\!\!\!\!
\Vert \bP_1Z_1(t)\Vert_{-2-\de}\le\fr{C(\ov d_1)} {(1+t)^{1+\de}}\Vert
Z(0)\Vert_{\beta}
\nonumber\\
\!\!\!\!\!\!\!\!\!\!\!\!\!\!\!\!\!\!&&\!\!\!\!\!\! +C(\ov
d_1)\int_0^t\fr1{(1+|t-s|)^{1+\de}}\left[\Vert Z(s)\Vert_{-\beta}
\int_s^{t_1}\Vert Z(\tau)\Vert^2_{-\beta}d\tau+ \Vert
Z(s)\Vert^2_{-\beta}\right]ds,\,\,\,0\le t\le t_1, \eeqn

Let us
introduce the {\it majorant} \be\la{maj} m(t):=
\sup_{s\in[0,t]}(1+s)^{1+\de}\Vert
Z(s)\Vert_{-\beta}~,~~~~~~~~~t\in [0,t_*). \ee
To estimate $d_{1}(t)$ by $m(t_1)$ we note that
\be\la{wen}
w(s)-v(t_{1})=w(s)-v(s)+v(s)-v(t_{1})=
\dot c(s)+\int_s^{t_1}\dot v(\tau)d\tau
\ee
by (\re{vw}).
Hence, (\ref {dd1}),
Lemma \ref{mod} and the definition (\re{maj}) imply
\begin{eqnarray}
\!\!\!\!\!\! \!\!\! \!\!\!
 |d_{1}(t)|\!\!\! &=&\!\!\!|\int_{t_{1}}^{t}(w(s)-v(t_{1}))ds|\leq
\int_{t}^{t_{1}}\left( |\dot{c}(s)|+\int_{s}^{t_{1}}|\dot{v}(\tau )|d\tau
\right)ds   \nonumber  \label{d1est} \\
&&  \nonumber \\
\!\!\! &\leq &\!\!\!Cm^{2}(t_{1})\int_{t}^{t_{1}}\left( \frac{1}{(1+s)^{2+2\de}}%
+\int_{s}^{t_{1}}\frac{d\tau }{(1+\tau )^{2+2\de}}\right) ds\leq
Cm^{2}(t_{1}),~~~~0\leq t\le t_{1}.
\end{eqnarray}
We can
replace in (\re{duhestr}) the constants $C(\ov d_1)$ by $C$ if $m(t_1)$ is
bounded for $t_1\ge 0$. In order to do this replacement, we reduce
the exit time.
 Let us denote by
$\ve$ a fixed positive number which we will specify below.
\begin{definition} $t_{*}'$ is the exit time
\be\la{t*'}
t_*'=\sup \{t\in[0,t_*):
m(s)\le \ve,~~0\le s\le t\}.
\ee
\end{definition}

Now (\re{duhestr}) implies that for $t_1<t_*'$ \beqn\la{duhestri}
\!\!\!\!\!\!\!\!\!\!\!\!\!\!\!\!\!\!&&\!\!\!\!\!\! \Vert
\bP_1Z_1(t)\Vert_{-2-\de}\le\fr{C} {(1+t)^{1+\de}}\Vert
Z(0)\Vert_{\beta}
\nonumber\\
\!\!\!\!\!\!\!\!\!\!\!\!\!\!\!\!\!\!&&\!\!\!\!\!\!
+C\int_0^t\fr1{(1+|t-s|)^{1+\de}}\left[\Vert Z(s)\Vert_{-\beta}
\int_s^{t_1}\Vert Z(\tau)\Vert^2_{-\beta}d\tau+ \Vert
Z(s)\Vert^2_{-\beta}\right]ds,\,\,\,0\le t\le t_1, \eeqn

%%%%%%%%%%%%%%%%%%%%%%%%%%%%%%%%%%%%%%%%%%%%%%%%%%%%

\subsection{Symplectic Orthogonality}

The following important bound  (\re{Z1P1est})
allows us to change the norm of $\bP_1Z_1(t)$ in the left hand side
 of
(\re{duhestri}) by the norm of $Z(t)$.

%%%%%%%%%%%%%%%%%%%%%%%%%%%%%%%%%%%%%%%%%%%%%%%%%%%%%%
\begin{lemma}\la{Z1P1Z1}
{\rm \ci[Lemma 13.1]{IKV09}} For sufficiently small $\ve>0$, we have for $t_1<t_*'$
\be\la{Z1P1est} \Vert Z(t)\Vert_{-2-\de}\le C\Vert \bP_1Z_1(t)\Vert_{-2-\de},
~~~~~~~~0\le t \le t_1, \ee where $C$ depends only on $\rho$ and $\ov v$.
\end{lemma}

%%%%%%%%%%%%%%%%%%%%%%%%%%%%%%%%%%%%%%%%%%%%%%%%%%%%%%%%%%%%%%%%%%%%%%%%%%%%%

\subsection{Decay of Transversal Component} Here we complete the proof of
Proposition \re{pdec}.
\\
{\it Step i)} We fix $\ve$, $0<\ve<r_{-\beta}(\ti v)$ and $t'_*=t'_*(\ve)$
for which Lemma \re{Z1P1Z1} holds.
Then the bound of type (\re{duhestri})
holds with
$\Vert \bP_1Z_1(t)\Vert_{-2-\de}$ in the left hand side replaced by
 $\Vert Z(t)\Vert_{-\beta}$~:
\beqn\la{duhestrih}
\!\!\!\!\!\!\!\!\!\!\!\!\!\!\!\!\!\!&&\!\!\!\!\!\! \Vert
Z(t)\Vert_{-\beta}\le\Vert
Z(t)\Vert_{-2-\de}\le C\Vert\bP_1
Z_1(t)\Vert_{-2-\de}     \le\fr{C} {(1+t)^{1+\de}}\Vert Z(0)\Vert_{\beta}
\nonumber\\
\!\!\!\!\!\!\!\!\!\!\!\!\!\!\!\!\!\!&&\!\!\!\!\!\!
+C\int_0^t\fr1{(1+|t-s|)^{1+\de}}\left[\Vert Z(s)\Vert_{-\beta}
\int_s^{t_1}\Vert Z(\tau)\Vert^2_{-\beta}d\tau+ \Vert
Z(s)\Vert^2_{-\beta}\right]ds,\,\,\,0\le t\le t_1 \eeqn for
$t_1<t_*'$. This implies an integral inequality
 for the majorant $m(t)$ introduced by (\re{maj}).
Namely, multiplying both sides of (\re{duhestrih}) by
$(1+t)^{1+\de}$, and taking the supremum in $t\in[0,t_1]$, we get
$$
m(t_1) \le C\Vert Z(0)\Vert_{\beta}+ C\sup_{t\in[0,t_1]}\ds
\int_0^t\fr{(1+t)^{1+\de}}{(1+|t-s|)^{1+\de}}\left[\fr{m(s)}{(1+s)^{1+\de}}
\int_s^{t_1}\fr{m^2(\tau)d\tau}{(1+\tau)^{2+2\de}}+\fr{m^2(s)}
{(1+s)^{2+2\de}}\right]ds
$$
for $t_1\le t_*'$. Taking into account that $m(t)$ is a monotone
increasing function, we get
\be\la{mest}
m(t_1)\le C\Vert Z(0)\Vert_{\beta}+C[m^3(t_1)+m^2(t_1)]I(t_1),
~~~~~~~~~~~~~t_1\le t_*'.
\ee
where
$$
I(t_1)= \sup_{t\in[0,t_1]}
\int_0^{t}\fr{(1+t)^{1+\de}}{(1+|t-s|)^{1+\de}}\left[\fr1{(1+s)^{1+\de}}
\int_s^{t_1}\fr{d\tau}{(1+\tau)^{2+2\de}}+\fr1{(1+s)^{2+2\de}}\right]ds
\le \ov I<\infty,~~~~t_1\ge0.
$$
Therefore, (\re{mest}) becomes
\be\la{m1est}
m(t_1)\le C\Vert Z(0)\Vert_{\beta}+C\ov I[m^3(t_1)+m^2(t_1)],~~~~ t_1<t_*'.
\ee
This inequality implies that $m(t_1)$
is bounded for $t_1<t_*'$, and moreover,
\be\la{m2est}
m(t_1)\le C_1\Vert Z(0)\Vert_{\beta},~~~~~~~~~t_1<t_*'\,,
\ee
since $m(0)=\Vert Z(0)\Vert_{\beta}$ is sufficiently small by
 (\re{closeZ}).

\medskip

\noindent
{\it Step ii)} The constant $C_1$ in the estimate
(\re{m2est}) does not depend on
$t_*$ and $t_*'$ by Lemma \re{Z1P1Z1}.
We choose $d_{\beta}$ in (\re{close}) so small that
$\Vert Z(0)\Vert_{\beta}<\ve/(2C_1)$. It is possible due to (\re{closeZ}).
Then the estimate (\re{m2est}) implies that $t'_*=t_*$ and therefore
 (\re{m2est}) holds for all $t_1<t_*$.
Then the bound (\re{d1est}) holds for all $t<t_*$. Therefore, (\re{m2est}) holds for
all $t_1<t_*$ and (\re{Zt}) holds as well. Finally, this implies that $t_*=\infty$,
hence also $t'_*=\infty$ and (\re{m2est}) holds for all $t_1>0$ if $d_{\beta}$ is small
enough. \hfill$\Box$

The transversal decay (\re{Zdec}) is proved.

 %%%%%%%%%%%%%%%%%%%%%%%%%%%%%%%%%%%%%%%%%%%%%%%%%%%%%%%%%%%%%%
\section{Soliton Asymptotics}
\setcounter{equation}{0}

Here we prove our main
Theorem \re{main} relying on
the decay (\re{Zdec}). First we will prove the asymptotics
(\re{qq}) for the vector components, and afterwards
the asymptotics (\re{S}) for the fields.

\medskip

{\bf Asymptotics for the vector components} From (\re{add})
we
have $\dot q=\dot b+\dot r$, and from (\re{reduced}), (\re{redN}),
(\re{AA}) it follows that $\dot r=-B_{v(t)}\langle\rho,a\rangle+B_{v(t)}\pi+
{\cal O} (\Vert
Z\Vert^2_{-\beta})$. Recall that $\beta=4+\de$, $0<\de<1/2$ Thus,
\be\la{dq}
\dot q=\dot b+\dot
r=v(t)+\dot c(t)-B_{v(t)}\langle\rho,a\rangle+B_{v(t)}\pi+{\cal O}
(\Vert Z\Vert^2_{-\beta}).
\ee
The equation
(\re{parameq}) and the estimates (\re{NZ}), (\re{Zdec}) imply that
\be\la{bv}
|\dot c(t)|+|\dot v(t)|\le \ds\fr {C_1(\rho,\ov
v,d_\beta)}{(1+t)^{2+2\de}}, ~~~~~~t\ge0.
\ee
Therefore, $c(t)=c_+
+\cO(t^{-1-2\de})$ and $v(t)=v_+ +\cO(t^{-1-2\de})$,
$t\to\infty$. Since $\Vert a\Vert_{-2-\de}$ and $|\pi|$ decay like $(1+t)^{-1-\de}$,
the estimate
(\re{Zdec}), and (\re{bv}), (\re{dq}) imply that
\be\la{qbQ}
\dot q(t)=v_++\cO(t^{-1-\de}).
\ee
Similarly,
\be\la{bt}
b(t)=c(t)+\ds\int_0^tv(s)ds=v_+t+a_++\cO(t^{-2\de}),
\ee
hence the
second part of (\re{qq}) follows:
\be\la{qbQ2}
q(t)=b(t)+r(t)=v_+t+a_++\cO(t^{-2\de}),
\ee
since
$r(t)=\cO(t^{-1-\de})$ by  (\re{Zdec}).

\medskip

{\bf Asymptotics for the fields} We apply the approach developed in \ci{IKSs}, see
also \ci{IKM,KKS}. For the field part of the solution, $F(t)=(E(x,t),A(x,t))$ let us
define the accompanying soliton field as $F_{\rm v(t)}(t)=(E_{\rm
v(t)}(x-q(t)),A_{\rm v(t)}(x-q(t)))$, where ${\rm v}(t):=\dot q(t)$. Then for the
difference $Z(t)=F(t)-F_{\rm v(t)}(t)$ we obtain easily the equation \ci{KKS}, Eq.
(2.5),
$$%\be\la{acc}
\dot Z(t)={\cal A}Z(t)-\dot{\rm v}\cdot\na_{\rm v}F_{{\rm v}(t)}(t),\,\,\,\,\,\,
{\cal A}(E,A)=(-\De A,-E).
$$%\ee
Then \be\la{eqacc} Z(t)=W^0(t)Z(0)-\int_0^tW^0(t-s)[\dot{\rm
v}(s)\cdot\na_{\rm v}F_{{\rm v}(s)}(s)]ds. \ee To obtain the
asymptotics (\re{S}) it suffices to prove that
$Z(t)=W^0(t)\Psi_++r_+(t)$ with some $\Psi_+\in{\cal F}$ and
$\Vert r_+(t)\Vert_{{\cal F}}={\cal O}(t^{-\de})$. This is
equivalent to \be\la{Sme} W^0(-t)Z(t)=\bPsi_++r_+'(t), \ee where
$\Vert r_+'(t)\Vert_\cF=\cO(t^{-\de})$ since $W^0(t)$ is a unitary
group in the Sobolev space $\cF$ by the energy conservation for
the free wave equation. Finally,
 (\re{Sme}) holds since
 (\re{eqacc}) implies that
\be\la{duhs} W^0(-t)Z(t)= Z(0)+\int_0^t W^0(-s)R(s)ds,\,\,\,\,\,R(s)=\dot{\rm
v}(s)\cdot\na_{\rm v}F_{{\rm v}(s)}(s), \ee where the integral  in the right hand
side of (\re{duhs}) converges in the Hilbert space $\cF$ with the rate
$\cO(t^{-\de})$. The latter holds since $\Vert W^0(-s)R(s)\Vert_\cF =\cO(s^{-1-\de})$
by the unitarity of $W^0(-s)$ and the decay rate $\Vert R(s)\Vert_\cF
=\cO(s^{-1-\de})$ which follows from the asymptotics for the vector components. More
precisely, differentiating the first
equation (\re{ph}) in $t$ and using the asymptotics
(\re{qbQ}), (\re{Zdec}) we obtain an estimate for $\dot{\rm v}(t)=\ddot q(t)$
providing the mentioned decay rate of $R(s)$.\hfill$\Box$

%%%%%%%%%%%%%%%%%%%%%%%%%%%%%%%%%%%%%%%%%%%%%%%%%%%%%%%%%%%%%%
%%%%%%%%%%%%%%%%%%%%%%%%%%%%%%%%%%%%%%%%%%%%%%%%%%%%%%%%%%%%%%

\setcounter{equation}{0}

\section{Solving the Linearized Equation}

In Sections 10-13 we prove Proposition \re{lindecay} in order to complete the
proof of the main result.
First, let us make a change of variables in the equation (\re{Avv})
to
simplify its structure. The equation (\re{Avv}) reads
\beqn\la{Avvp}
\dot e = & v\cdot\na e-\De a+\Pi_s(r\cdot\na\rho v-\rho
B_v(\pi-\langle\rho,a\rangle)), &
\dot a = -e+v\cdot\na a,\nonumber\\
\dot r  = & B_v(\pi-\langle\rho,a\rangle), & \dot\pi = \langle\rho,\na(v\cdot
a)\rangle-\langle r\cdot\na\rho,\na(v\cdot A_v)\rangle.
\eeqn
Put
$\varphi=\pi-\langle\rho,a\rangle$. Then $\pi=\varphi+\langle\rho,a\rangle$. If we
prove a decay of $\varphi$ and $a$, then $\pi$ has the corresponding decay as well.
Further, $\dot\varphi=\dot\pi-\langle\rho,\dot
a\rangle=\dot\pi-\langle\rho,-e+v\cdot\na a\rangle=\langle\rho,e\rangle+
\langle\rho,\na(v\cdot a)-(v\cdot\na)a\rangle-\langle
r\cdot\na\rho,\na(v\cdot A_v)\rangle$ by the last equation of (\re{Avvp}). Thus, the
system (\re{Avvp}) is equivalent to the following system,
\beqn\la{Avvpi}
\dot e = &
v\cdot\na e-\De a+\Pi_s(r\cdot\na\rho v-\rho B_v\varphi), &
\dot a = -e+v\cdot\na a,\nonumber\\
\dot r  = & B_v\varphi, & \dot\varphi = \langle\rho,e\rangle+
\langle\rho,v\we(\na\we a)\rangle-\langle
r\cdot\na\rho,\na(v\cdot A_v)\rangle.
\eeqn
For the last equation we have applied the identity $\na(v\cdot a)-(v\cdot\na)a=
v\we(\na\we a)$.
Denote by the same letter $A$ the operator
\beqn\la{Api}
A\left( \ba{c} e \\ a \\ r \\ \varphi \ea \right):=\left( \ba{l}
v\cdot\na e-\De a+\Pi_s(r\cdot\na\rho v-\rho B_v\varphi) \\
-e+v\cdot\na a \\
B_v\varphi, \\
\langle\rho,e\rangle+\langle\rho,v\we(\na\we a)\rangle-
\langle r\cdot\na\rho,\na(v\cdot A_v)\rangle
\ea\right).
\eeqn
Below we prove the decay for the solution $X=(e,a,r,\vp)$ to the equation
\be\la{AAA}
\dot X(t)=AX(t).
\ee
So, now we construct and study the resolvent of $A$.

Let us apply the Laplace transform
\be\la{FL}
\Lam X=\ti
X(\lam)=\int_0^\infty e^{-\lam t}X(t)dt,~~~~~~~\Re\lam>0 \ee to
(\re{Avv}). The integral converges in ${\cal E}$, since $\Vert X(t)\Vert_{\cal E}$
is bounded by Proposition \re{lindecay}, i). The
analyticity of $\ti X(\lam)$ and Paley-Wiener arguments should
provide the existence of a $\cE$-valued distribution $X(t)=(\Psi(t),\Pi(t),Q(t),P(t))$,
$t\in\R$, with a support in $[0,\infty)$. Formally,
\be\la{FLr}
\Lam^{-1}\ti X=X(t)=\fr1{2\pi}\int_\R e^{i\om t}\ti X(i\om+0)d\om,
~~~~~~~~t\in\R. \ee To prove the decay (\re{frozenest}) we have to
study the smoothness of $\ti X(i\om+0)$ at $\om\in\R$. After
the Laplace transform the equation (\re{Avv}) becomes
\be\la{FLe}
\lam\ti X(\lam)=A\ti X(\lam)+X_0,\,\,\,\Re\lam>0.
\ee

%%%%%%%%%%%%%%%%%%%%%%%%%%%%%%%%%%%%%%%%%%%%%%%%%%%%%%%%%%%%%%%%%%%%%

%%%\subsection{Constructing the Resolvent}

To justify the representation (\re{FLr}),
we
construct the resolvent as a bounded operator in
${\cal E}$
 for $\Re\lam>0$.
We shall write $(e(y),a(y), r, \varphi)$ instead of $(\ti e(y,\lam),\ti a(y,\lam),\ti
r(\lam),\ti{\varphi}(\lam))$ to simplify the notations. Then (\re{FLe}) reads
\be\la{eq1}
\left.\ba{l} v\cdot\na e-\De a+\Pi_s(r\cdot\na\rho v-\rho
B_v\varphi)-\lam e = -e_0,\,\,\,\,\,\,\,\,-e+v\cdot\na a-\lam a = -a_0
\\
\\
B_v\varphi-\lam r = -r_0,\,\,\,\,\,\,\,\,\langle\rho,e\rangle+
\langle\rho,v\we(\na\we a)\rangle-\langle r\cdot\na\rho,\na(v\cdot
A_v)\rangle-\lam\varphi = -\varphi_0 \ea\right|
\ee
{\it Step i)} Let us consider the
first two equations. After Fourier transform
 they become
\be\la{F1} -i(kv)\hat e+k^2\hat a-\hat\Pi_s(i(kr)\hat\rho v+\hat\rho
B_v\varphi)-\lam\hat e=-\hat e_0,\,\,\,-\hat e-i(kv)\hat a-\lam\hat a=-\hat a_0 \ee
>From the last equation we have $\hat e=-(\lam+i(kv))\hat a+\hat a_0$. Substitute to
the first equation of (\re{F1}) and obtain \be\la{hata} \hat a=\fr1{\hat
D}((\lam+ikv)\hat a_0-\hat
e_0+\hat\Pi),\,\,\,\hat\Pi:=\hat\rho\hat\Pi_s(i(kr)v+B_v\varphi)), \ee where
\be\la{Dlam} \hat D=\hat D(\lam)=k^2+(\lam+ikv)^2. \ee It is easy to see that $\hat
D(\lam)\ne 0\,\,\,{\rm for}\,\,\,\Re\lam>0$. Finally, \be\la{hate} \hat e=\fr{k^2\hat
a_0+(\lam+ikv)\hat e_0-(\lam+ikv)\hat\Pi}{\hat D}. \ee

\medskip

\noindent{\it Step ii)}
Let us proceed to the fourth
equations of (\re{eq1}). The equation reads
$$
\langle\rho,e\rangle+\langle\rho,v\we(\na\we a)\rangle-\langle
r\cdot\na\rho,\na(v\cdot A_v)\rangle-\lam\varphi=-\varphi_0.
$$
>From now on we use the system of coordinates in $x$-space in which $v=(|v|,0,0)$,
hence $vk=|v|k_1$. By (\re{hate}) and a straightforward computation we obtain
$$
\langle\rho,e\rangle=\Phi-C_1r+F_1\varphi,
$$
where \be\la{Phioflam} \Phi=\Phi(\lam,e_0,a_0):=\int(k^2\hat a_0+(\lam+ikv)\hat
e_0)\ov{\hat\rho}/\hat D\,dk \ee and $C_1$, $F_1$ are the following diagonal
$3\times3$-matrices: \be\la{C1F1} C_1(\lam)=\left( \ba{lll}
c_{11}(\lam) & 0 & 0\\
0 & c_{12}(\lam) & 0\\
0 & 0 & c_{13}(\lam) \ea\right),\,\,\,F_1(\lam)=\left( \ba{lll}
f_{11}(\lam) & 0 & 0\\
0 & f_{12}(\lam) & 0\\
0 & 0 & f_{13}(\lam) \ea\right), \ee
\be\la{c1}c_{11}(\lam)=i|v|\int\fr{k_1(\lam+ik_1|v|)|\hat\rho|^2}{\hat
D(\lam)}(1-\fr{k_1^2}{k^2})dk,\,\,
c_{1j}(\lam)=-i|v|\int\fr{k_1k_j^2(\lam+ik_1|v|)|\hat\rho|^2}{k^2\hat
D(\lam)}dk,\,\,j=2,3, \ee \be\la{f123}
f_{11}(\lam)=\nu^3\int\,\fr{(\lam+ik_1|v|)|\hat\rho|^2}{\hat
D(\lam)}(\fr{k_1^2}{k^2}-1)dk,\,\,f_{1j}(\lam)=
\nu\int\,\fr{(\lam+ik_1|v|)|\hat\rho|^2}{\hat
D(\lam)}(\fr{k_j^2}{k^2}-1)dk,\,\,\,j=2,3; \ee recall that $\nu=\sqrt{1-v^2}$. By the
change of variables $k_2\mapsto k_3$ we obtain that $c_{12}=c_{13}$. Moreover,
$c_{11}+c_{12}+c_{13}=0$ and thus, $c_{11}=-c_{12}-c_{13}=-2c_{12}$. The matrix
$C_1(\lam)$ simplifies to \be\la{CG} C_1(\lam)=\left( \ba{lll}
c_1(\lam) & 0 & 0\\
0 & c_{12}(\lam) & 0\\
0 & 0 & c_{12}(\lam)
\ea\right),\,\,c_1(\lam):=-2c_{12}(\lam),\,\,c_{12}(\lam)=
-\fr{i|v|}2\int\fr{k_1(\lam+ik_1|v|)|\hat\rho|^2}{\hat
D(\lam)} (1-\fr{k_1^2}{k^2})dk. \ee Similarly, $f_{12}=f_{13}$ and
$$
f_{12}(\lam)=\nu\int\,\fr{(\lam+ik_1|v|)|\hat\rho|^2}{\hat
D(\lam)}(\fr{k_2^2+k_3^2}{2k^2}-1)dk=\nu\int\,\fr{(\lam+ik_1|v|)|\hat\rho|^2}{\hat
D(\lam)}(\fr{k^2-k_1^2}{2k^2}-1)dk=
$$
\be\la{fup1} -\fr{\nu}2\int\,\fr{(\lam+ik_1|v|)|\hat\rho|^2}{\hat
D(\lam)}(1+\fr{k_1^2}{k^2})dk \ee  Further, $\langle\rho,v\we(\na\we
a)\rangle=\Psi-C_2r+F_2\vp$, where \be\la{Psilam}
\Psi=\Psi(\lam,e_0,a_0):=\int\,dk\,v\we(-ik\we\fr{(\lam+ikv)\hat a_0-\hat e_0}{\hat
D(\lam)})\ov{\hat\rho} \ee and \be\la{C2F2lam} C_2(\lam)=\left( \ba{ccc}
0 & 0 & 0 \\
0 & c_{22}(\lam) & 0 \\
0 & 0 & c_{23}(\lam) \ea \right),\,\,\,F_2(\lam)=\left( \ba{ccc}
0 & 0 & 0 \\
0 & f_{22}(\lam) & 0 \\
0 & 0 & f_{23}(\lam) \ea \right), \ee where \be\la{c2f2lam}
c_{2j}(\lam)=-v^2\int\,dk\fr{|\hat\rho|^2k_j^2}{\hat
D(\lam)},\,\,\,f_{2j}(\lam)=i\nu|v|\int\,dk\fr{|\hat\rho|^2k_1}{\hat
D(\lam)},\,\,\,j=2,3. \ee \br{\rm Note that \be\la{philam} \Phi(\lam)=\Lam\langle
W^1(t)(e_0,a_0),\rho\rangle, \ee where $W^1(t)$ is the first component of the
dynamical group $W(t)$ defined below by (\re{184st}). Similarly, \be\la{Psioflam}
\Psi(\lam)=\Lam\langle v\we(\na\we W^2(t)(e_0,a_0)),\rho\rangle, \ee where $W^2(t)$
is the second component of the same dynamical group.} \er

\smallskip

\noindent Further, $c_{22}=c_{23}$, $f_{22}=f_{23}$ and \be\la{cup2lam}
c_{22}(\lam)=-\fr{v^2}2\int\,dk\fr{|\hat\rho|^2(k_2^2+k_3^2)}{\hat
D(\lam)}=-\fr{v^2}2\int\,dk\fr{|\hat\rho|^2(k^2-k_1^2)}{\hat D(\lam)}. \ee At last,
$\langle r\cdot\na\rho,\na(v\cdot A_v)\rangle=Gr$, where \be\la{G} G=\left( \ba{lll}
g_1 & 0 & 0\\
0 & g_2 & 0\\
0 & 0 & g_3
\ea\right),\,\,g_j=v^2\int\fr{(k^2-k_1^2)k_j^2|\hat\rho|^2}{k^2(k^2-k_1^2v^2)}dk
,\,\,\,j=1,2,3.
\ee Again, \be\la{gg23}g_2=g_3\,\,\,{\rm and\,\, we\,\, set}\,\,\,g:=g_2=g_3.\ee Put
\be\la{CFlam}C(\lam)=C_1(\lam)+C_2(\lam),\,\,\,F(\lam)=F_1(\lam)+F_2(\lam).\ee In
detail, by (\re{C1F1}), (\re{CG}), and (\re{C2F2lam}) \be\la{Cdet} C(\lam)=\left(
\ba{lll} c_1(\lam) & 0 & 0
\\ 0 & c(\lam) & 0
\\ 0 & 0 & c(\lam)\ea \right),\,\,\,c(\lam):=c_{12}(\lam)+c_{22}(\lam), \ee

\be\la{Fdet} F(\lam)=\left( \ba{lll} f_1(\lam) & 0 & 0 \\ 0 & f(\lam) & 0
\\ 0 & 0 & f(\lam)\ea \right),\,\,\,f_1(\lam):=f_{11}(\lam),\,\,\,
f(\lam):=f_{12}(\lam)+f_{22}(\lam).\ee Finally, the fourth equation becomes
$(C(\lam)+G)r+(\lam E-F(\lam))\varphi=\varphi_0+\Phi(\lam)+\Psi(\lam)$. We write this
equation and the third equation of (\re{eq1}) together in the form
\be\la{vec12} M(\lam)\left( \ba{l} r \\
\varphi \ea\right)=\left( \ba{l} r_0 \\ \varphi_0+\Phi(\lam)+\Psi(\lam)
\ea\right),\,\,\,{\rm where}\,\,\,M(\lam)=\left( \ba{ll}
\lam E & -B_v\\
C(\lam)+G & \lam E-F(\lam)
\ea\right).
\ee
Assume for a moment that the matrix
$M(\lam)$ is invertible for $\Re\lam>0$
 (see below).
Then \be\la{QP1} \left( \ba{c} r \\ \varphi \ea \right)=M^{-1}(\lam)\left( \ba{c} r_0
\\ \varphi_0+\Phi(\lam)+\Psi(\lam) \ea \right),~~~~~~~~~\Re\lam>0. \ee Formulas
(\re{hata}), (\re{hate}), and (\re{QP1}) give the expression of the resolvent
 $R(\lam)=(A-\lam)^{-1}$, $\Re\lam>0$, in Fourier representation.

\medskip

Further, the operator $D(\lam)$
defined in Fourier space as
multiplication by the symbol
(\re{Dlam}),
% is the same as in the case of wave
%equation studied in \ci{IKV09}. Then
%by \ci[Lemmas 7.3 and 7.4]{IKV09} the operator $D(\lam)$
 is invertible in $L^2(\R^3)$
for $\Re\lam>0$ and its fundamental solution
$g_\lam(y)$
exponentially decays as $|y|\to\infty$.

\bl \la{cac} i) The distribution $g_\lam(\cdot)$
admits an analytic continuation in $\lam$ from the domain
 $\Re\lam>0$ to the entire complex plane $\C$.
\\
ii) The matrix function $M(\lam)$ ($M^{-1}(\lam)$) admits an analytic (respectively,
meromorphic) continuation
in the parameter $\lam$
from the domain
 $\Re\lam>0$
to the entire complex plane. \el

\pru The fundamental solution $g_\lam(y)$ is
 given by

\be\la{glam} g_{\lam}(y)=\fr{e^{-\ka|\ti y|-\ka_1\ti y_1}}{4\pi|\ti y|},
\,\,\,\ti y:=(\ga y_1,y_2,y_3),
\ee
where
\be\la{gammakappa}
\ga:=1/\sqrt{1-v^2},\,\,\,\ka=\ga\lam,\,\,\,\ka_1:=|v|\ka.
\ee
Thus, the statement i)
follows from the formulas (\re{gammakappa}), (\re{glam}). To prove the statement ii)
we first need to show, according to (\re{vec12}) that the matrices $C(\lam)$ and
$F(\lam)$ admit an analytic continuation to the entire complex plane. Consider the
matrix $C(\lam)$, for $F(\lam)$ the argument is similar.  The analytic continuation
of $C(\lam)$ then exists by the expression of type (\re{C1x})
for the entries of the matrix $C(\lam)$,
and the statement i)
of the present lemma since the function $\rho(x)$
is compactly supported by
(\re{ro}). The inverse matrix $M^{-1}(\lam)$ is then meromorphic, since it is well
defined for large $|\lam|$ by (\re{Minvlam})--(\re{L2122}) and Corollary B.3.\hfill
$\Box$

%%%%%%%%%%%%%%%%%%%%%%%%%%%%%%%%%%%%%%%%%%%%%%%%%%%%%%%%%%%%%%%%%%%%%%%%%
\setcounter{equation}{0}
\section{Regularity in Continuous Spectrum}

By Lemma \re{cac}, the
limit matrix \be\la{M} M(i\om):=M(i\om+0)=\left( \ba{ll}
i\om E & -B_v \\
C(i\om+0)+G & i\om E-F(i\om+0)
\ea
\right), ~~~~~~~~~~\om\in\R,
\ee
exists, and its entries are analytic
functions of $\om\in\R$.
Recall that the point $\lam=0$ belongs to the discrete spectrum of
the operator $A$ by Lemma \re{ceig} i), hence
$M(i\om+0)$ (probably) is also not invertible at $\om=0$.
%%%%%%%%%%%%%%%%%%%%%%%%%%%%%%%%%%%%%%%%%%
\bp\la{regi}
The matrix $M^{-1}(i\om)$ is analytic in $\om\in\R\setminus\{0\}$.
\ep
%%%%%%%%%%%%%%%%%%%%%%%%%%%%%%%%%%%%%%%%%
{\bf Proof } It suffices to prove that the limit matrix
 $M(i\om):=M(i\om+0)$ is invertible for
$\om\ne 0$, $\om\in\R$ if $\rho$ satisfies the Wiener condition (\re{W}), and
$|v|<1$. Since $v=(|v|,0,0)$,  the matrix $B_v$ is also diagonal:
\be\la{Bv}
B_v:=\nu(E-v\otimes v)=
\left(
\ba{ccc}
\nu^3&0&0\\
0&\nu&0\\
0&0&\nu \ea \right). \ee By (\re{vec12}), (\re{gg23}), (\re{Cdet}), (\re{Fdet}),
(\re{Bv}), for $\om\in\R$,
$$
{\rm det}\,M(i\om)={\rm det}\,\left(
\ba{ll}
i\om E  & -B_v \\
C(i\om)+G & i\om E-F(i\om)
\ea
\right)=d_1d^2,
$$
where
\be\la{detM1}
d_1=-\om^2-i\om f_1(i\om)+\nu^3(c_1(i\om)+g_1),
\ee
\be\la{detM2}
d=-\om^2-i\om f(i\om)+\nu(c(i\om)+g).
\ee
The formula for the determinant
is obvious since all of the matrices $C$, $F$, $G$, and $B_v$ are diagonal.
Then for $|\om|>0$ the invertibility  of $M(i\om)$
follows from  (\re{detM1}), (\re{detM2})
by the following lemma.

\bl\la{lW}
If (\re{W}) holds and $|\om|>0$, then the imaginary parts of $d_1$ and $d$
are positive: $\Im d_1>0$, $\Im d>0$.
\el
\pru Let $\om>0$, the case $\om<0$ is
similar. Note that $\Im d_1=-\om\Re f_1(i\om+0)+\nu^3\Im\,c_1(i\om+0)$. For
$\ve>0$ we have
\be\la{Hjjlim}
f_{1}(i\om+\ve)=\nu^3\int\fr{(i\om+\ve+ik_1|v|)|\hat\rho(k)|^2}{\hat
D(i\om+\ve,k)}(\fr{k_1^2}{k^2}-1)dk.
\ee
By Sokhotsky-Plemelj formula for
$C^1$-functions, (\ci{Vai}, Chapter VII, formula (58)),
\be\la{ImHjj}
\Re f_{1}(i\om+0)=\pi\nu^3\int_{T_{\om}}\fr{(\om+k_1|v|)|\hat\rho(k)|^2}
{|\na\hat D(i\om,k)|}(\fr{k_1^2}{k^2}-1)dS,
\ee
where
$$
T_{\om}=\{k:k^2-(\om+k_1|v|)^2=0\}
$$
is the ellipsoid on which
$\hat D(i\om,k)=0$.
Similarly,
\be\la{Rec1}
\Re\,c_{1}(i\om+0)=\pi|v|\int_{T_{\om}}\fr{k_1(\om+k_1|v|)|\hat\rho(k)|^2} {|\na\hat
D(i\om,k)|}(1-\fr{k_1^2}{k^2})dS.
\ee
Then
$$
\Im d_1=\nu^3\pi\int_{T_{\om}}\fr{(\om+k_1|v|)^2|\hat\rho(k)|^2} {|\na\hat
D(i\om,k)|}(1-\fr{k_1^2}{k^2})dS>0
$$
by the Wiener condition (\re{W}).
Further,
$$
\Im d=-\om\Re f(i\om+0)+\nu\Im\,c(i\om+0)=
$$
$$
-\om\Re f_{12}(i\om+0)+\nu\Im\,c_{12}(i\om+0)-\om\Re
f_{22}(i\om+0)+\nu\Im\,c_{22}(i\om+0).
$$
By
(\re{fup1}) we have
$$
f_{12}(i\om+\ve)=-\fr{\nu}2\int\fr{(i\om+\ve+ik_1|v|)|\hat\rho(k)|^2}{\hat
D(i\om+\ve,k)}(1+\fr{k_1^2}{k^2})dk.
$$
Then
$$
\Re f_{12}(i\om+0)=-\fr{\pi\nu}2\int_{T_{\om}}\fr{(\om+k_1|v|)|\hat\rho(k)|^2}
{|\na\hat D(i\om,k)|}(1+\fr{k_1^2}{k^2})dS=
$$
$$
-\fr{\pi\nu}2\int_{T_{\om}}\fr{(\om+k_1|v|)|\hat\rho(k)|^2}
{|\na\hat D(i\om,k)|}\fr{(\om+k_1|v|)^2+k_1^2}{k^2}dS,
$$
since $k^2=(\om+k_1|v|)^2$ on $T_\om$. By (\re{CG}) we obtain that
$$
c_{12}(i\om+\ve)=\fr{|v|}2\int\fr{k_1(\om-i\ve+k_1|v|)|\hat\rho(k)|^2}{\hat
D(i\om+\ve,k)}(1-\fr{k_1^2}{k^2})dk.
$$
Then
$$
\Im\,c_{12}(i\om+0)=-\fr{\pi|v|}2\int_{T_{\om}}\fr{k_1(\om+k_1|v|)|\hat\rho(k)|^2}
{|\na\hat D(i\om,k)|}(1-\fr{k_1^2}{k^2})dS=
$$
$$
-\fr{\pi|v|}2\int_{T_{\om}}\fr{k_1(\om+k_1|v|)|\hat\rho(k)|^2}
{|\na\hat D(i\om,k)|}\fr{(\om+k_1|v|)^2-k_1^2}{k^2}dS
$$
and \be\la{ImDup1}
-\om\Re\,f_{12}+\nu\Im\,c_{12}=\fr{\nu\pi}2\int_{T_{\om}}
\fr{|\hat\rho(k)|^2(\om^2+k_1^2(1-v^2))}
{|\na\hat D(i\om,k)|}dk. \ee Further, by (\re{c2f2lam})
$$
\Re\,f_{22}(i\om+0)=\pi\nu|v|\int_{T_\om}dS\fr{|\hat\rho(k)|^2k_1}{|\na\hat
D(i\om,k)|},
$$
by (\re{cup2lam})
$$
\Im\,c_{22}(i\om+0)=\fr{\pi
v^2}2\int_{T_\om}dS\fr{|\hat\rho(k)|^2(k^2-k_1^2)}{|\na\hat D(i\om,k)|},
$$
thus,
$$
-\om\Re
f_{22}(i\om+0)+\nu\Im\,c_{22}(i\om+0)=\fr{\pi\nu}2\int_{T_\om}dS
\fr{|\hat\rho(k)|^2(v^2((\om+k_1|v|)^2-k_1^2)-2\om
k_1|v|)}{|\na\hat D(i\om,k)|}.
$$
Finally, combining with (\re{ImDup1}) we obtain
$$
\Im\,{\hat
d}=\fr{\pi\nu}2\int_{T_\om}dS\fr{|\hat\rho(k)|^2((k_1(v^2-1)+\om|v|)^2+\om^2)}{|\na\hat
D(i\om,k)|}>0
$$
by the Wiener condition. This completes the proofs of the lemma and
the Proposition \re{regi}.
\hfill $\Box$

\br
{\rm
The proof of the  Lemma \re{lW} is the unique point
in the paper where
the Wiener condition is indispensable.}
\er

%%%%%%%%%%%%%%%%%%%%%%%%%%%%%%%%%%%%%%%%%%%%%%%%%%%%%%%%%%%
\setcounter{equation}{0}
\section{Time Decay of the Vector Components}

Let us prove the decay (\re{frozenest}) for the vector components $r(t)$ and
$\phi(t)$ of the solution $e^{At}X_0$. Formula (\re{QP1}) expresses the Laplace
transforms $\ti r(\lam),\ti \varphi(\lam)$. Hence, the components are given by the
Fourier integral
\be\la{QP1i}
\left( \ba{c} r(t) \\ \vp(t) \ea \right)=\ds\fr
1{2\pi}\int e^{i\om t}M^{-1}(i\om)\left( \ba{c} r_0 \\ \vp_0+\Phi(i\om)+\Psi(i\om)
\ea \right)d\om.
\ee
Recall that in Proposition \re{lindecay} we assume that
\be\la{ortog} X_0\in{\cal Z}_v\cap{\cal E}_\beta,\,\,\,\beta=4+\de,\,\,\,0<\de<1/2.
\ee
\bt\la{171}
The functions $r(t)$, $\vp(t)$ are continuous  for $t\ge 0$,
and \be\la{decQP}
|r(t)|+|\vp(t)|\le \ds\fr {C(\rho, \ti v,d_\beta)}{(1+|t|)^{1+\de}}\Vert
X_0\Vert_\beta, ~~~~t\ge 0.
\ee
\et
\pru The Proposition \re{regi} alone is not
sufficient for the proof of the convergence and decay of the integral. Namely, we
need an additional information about a regularity of the matrix $L(i\om)$ and of
$\Phi(i\om)+\Psi(i\om)$.
 Let us split the Fourier integral
(\re{QP1i}) into two terms using the partition of unity
$\zeta_1(\om)+\zeta_2(\om)=1$, $\om\in\R$: $$ \left( \ba{c}
r(t) \\
\vp(t)
\ea \right)=\ds\fr 1{2\pi}\int e^{i\om t}(\zeta_1(\om)+\zeta_2(\om))
\left( \ba{c}
\ti r(i\om) \\
\ti \vp(i\om)
\ea \right)
d\om =
$$
\be\la{QP1i3} \left( \ba{c}
r_1(t) \\
\vp_1(t)
\ea \right)+\left( \ba{c}
r_2(t) \\
\vp_2(t) \ea \right)=I_1(t)+I_2(t), \ee where the functions $\zeta_k(\om)\in
C^\infty(\R)$ are supported by \be\la{zsup} \supp
\zeta_1\subset\{\om\in\R:|\om|<r+1\},\,\,\,\supp\zeta_2\subset\{\om\in\R:|\om|>r\},\ee
where $r$ is introduced below in Lemma \re{162}. We prove the decay (\re{decQP}) for
$(r_1,\vp_1)$ and $(r_2,\vp_2)$ in Propositions \re{r1p1} and \re{r2p2} respectively.

\bp\la{r2p2}
The function $I_2(t)$ is continuous for $t\ge 0$, and
\be\la{I2dec}
|I_2(t)|\le C(\rho,\ti v)(1+|t|)^{-3-\de}\Vert
X_0\Vert_\beta.
\ee
\ep
\pru First, we need the asymptotic behavior of $M^{-1}(\lam )$ at
infinity. Let us recall that $M^{-1}(\lam)$ was originally defined
for $\Re\lam >0,$ but
it admits a meromorphic continuation to the entire complex plane $\C$
(see Lemma
\ref{cac}).

\bl \la{162} There exist a matrix $R_{0}$ and a matrix-function $R_{1}(\om )$,  such
that \be\la{Minf} M^{-1}(i\om)=\fr {R_0}\om +R_1(\om ),~~~|\om|>r>0, ~~~~~~~~~\om \in
\R, \ee where, for every $k=0,1,2,...,$ \be\label{min} |\pa_\om^k R_1(\om )|\leq
\fr{C_k}{|\om |^2}, ~~~~~~~~~~~~|\om|>r>0,~~~~~~~~~\om \in \R, \ee $r$ is
sufficiently large. \el \pru The statement follows from the explicit formulas
(\re{detMD1D}), (\re{Minvlam})--(\re{L2122}) for the inverse matrix $M^{-1}(i\om)$
and from Lemma B.2.\hfill $\Box$

\bigskip

Further, (\re{QP1i}) implies that
$$
I_2(t)=\fr 1{2\pi}\int e^{i\om t}\zeta_2(\om)
M^{-1}(i\om)
\left[\left(\ba{c}
r_0 \\
\vp_0
\ea \right)+
\left( \ba{c}
0 \\
\Phi(i\om)+\Psi(i\om)
\ea \right)
\right]d\om
$$
\be\la{long}
=s(t)\left(
\ba{c}
r_0 \\ \vp_0
\ea
\right)+s*\left(
\ba{c}
0 \\ f+\psi
\ea
\right),
\ee
where (see(\re{FLr}))
$$
s(t):=\Lam^{-1}\left[\zeta_2(\om)
 M^{-1}(i\om)\right],
$$
and
\be\la{fphi}
f(t):=\Lam^{-1}\Phi(i\om)=\langle W^1(t)(e_0,a_0),\rho\rangle,\,\,\,
\psi(t):=\Lam^{-1}\Psi(i\om)=\langle v\we(\na\we W^1(t)(e_0,a_0)),\rho\rangle
\ee
since $\Phi$, $\Psi$ are given by (\re{Phioflam}), (\re{Psilam}),
(\re{philam}), (\re{Psioflam}).
Recall that $(e_0,a_0,r_0,\pi_0)=X_0\in \cF_\beta$ with
$\beta=4+\de$
where $\de>0$ under conditions of Proposition \re{lindecay} and Theorem \re{main}.
Hence, applying Lemma \re{lemma} (see below) with $\al=4+\de$, we
obtain that
\be\la{fdec}
|f(t)|+|\psi(t)|\le C(\rho,\ti v)(1+t)^{-3-\de}\Vert X_0\Vert_\beta.
\ee
On the other hand,
(\re{Minf})-(\re{min}) imply that
$$
|s(t)|={\cal O}(t^{-N}),\,\,\,|t|\to\infty,\,\,\,\forall\,N>0.
$$
Hence, all the terms in (\re{long}) are continuous
for $t\ge 0$ and decay like $Ct^{-3-\de}\Vert X_0\Vert_\beta$.
\hfill $\Box$

\medskip

\noindent Now let us prove the decay for $I_1(t)$. In this case the proof will rely
substantially on the symplectic orthogonality conditions. Namely, (\re{ortog})
implies that \be\la{ortogg} \Om(X_0,\tau_j)=0,\,\,\,\,\,\,j=1,\ldots,6. \ee
\bp\la{r1p1}
The function $I_1(t)$ is continuous for $t\ge0$, and
\be\la{PQdec} |I_1(t)|\le C(\rho,\ti v)(1+t)^{-1-\de}\Vert
X_0\Vert_\beta,\,\,\,t\ge0.
\ee
\ep
\pru First, let us calculate  the Fourier transforms
$\ti r_1(i\om)$ and $\ti\vp_1(i\om)$.
\bl\la{PiomQiom}
The matrix  $M^{-1}(i\om)$ can be represented as follows,
\be\la{calLom} M^{-1}(i\om)=\left( \ba{cc}
\fr{\ts 1}{\ds\om}{\cal L}_{11} & \fr{\ts 1}{\ds\om^2}{\cal L}_{12} \\
{\cal L}_{21} & \,\,\,\fr{\ts 1}{\ds\om}{\cal L}_{22}
\ea
\right),
\ee
where ${\cal L}_{ij}(\om)$, $i,j=1,2$ are smooth diagonal
$3\times3$-matrices,
${\cal L}_{ij}(\om)\in C^{\infty}(-r-1,r+1)$.
Moreover,
\be\la{m11m12}
{\cal L}_{11}=i{\cal L}_{12}B_v^{-1}+i{\cal L}_3,
\ee
where ${\cal L}_3$ is defined by (\re{calL3}), ${\cal L}_3$ is a smooth diagonal
$3\times3$-matrix, ${\cal L}_{3}(\om)\in C^{\infty}(-r-1,r+1)$.
\el
For proof see Appendix B.
Then the vector components are given by
\be\la{tiQiom}
\ti r(i\om)=\fr{1}{\om}{\cal L}_{11}(\om)r_0+\fr{1}{\om^2}{\cal L}_{12}(\om)
(\vp_0+\Phi(i\om)+\Psi(i\om)),
\ee
\be\la{tiPiom}
\ti\vp(i\om)={\cal L}_{21}(\om)r_0+\fr{1}{\om}{\cal L}_{22}(\om)(\vp_0+\Phi(i\om)+
\Psi(i\om)).
\ee
Next we calculate the symplectic orthogonality conditions (\re{ortogg}).
\bl\la{sort123} The symplectic orthogonality conditions (\re{ortogg}) read
\be\la{ort123}
\vp_0+\Phi(0)+\Psi(0)=0\,\,\,{\rm and}\,\,\,
(B_v^{-1}+{\cal L}_{12}^{-1}(0){\cal L}_3(0))r_0+\Phi'(0)+\Psi'(0)=0.
\ee
\el
For proof see Appendix C.

\medskip
Now we can prove Proposition \re{r1p1}.

\noindent
Step i) Let us prove (\re{PQdec}) for $\vp_1(t)$ relying on the representation
(\re{tiPiom}). Namely, (\re{QP1i3}) and (\re{tiPiom}) imply
$$
\vp_1(t)=\Lam^{-1}\zeta_1(\om){\cal L}_{21}(\om)r_0+
\Lam^{-1}\zeta_1(\om){\cal L}_{22}(\om)\fr{\vp_0+\Phi(i\om)+\Psi(i\om)}{\om}=
\vp_1'(t)+\vp_1''(t).
$$
The first term $\vp_1'(t)$ decays like $Ct^{-\infty}\Vert X_0\Vert_\beta$ by Lemma
\re{PiomQiom}. The second term admit the convolution representation
$\vp_1''(t)=\Lam^{-1}\zeta_1{\cal L}_{22}*g(t)$, where
$$
g(t):=\Lam^{-1}\fr{\vp_0+\Phi(i\om)+\Psi(i\om)}{\om}.
$$
Now we use the symplectic orthogonality conditions (\re{ort123}) and obtain
$$
g(t)=\Lam^{-1}\fr{\Phi(i\om)+\Psi(i\om)-\Phi(0)-\Psi(0)}{\om}=
i\int\limits^{t}_{\infty}(f(s)+\psi(s))ds.
$$
Finally, $g(t)$ decays like $Ct^{-2-\de}\Vert X_0\Vert_\beta$ for $t\ge0$ by
(\re{fdec}), hence $\vp_1''(t)$ decays like $Ct^{-2-\de}\Vert X_0\Vert_\beta$
for $t\ge0$.

\medskip

\noindent Step ii). Now let us prove (\re{PQdec}) for $r_1(t)$. By (\re{tiQiom}),
(\re{m11m12}), and the symplectic orthogonality conditions (\re{ort123}),
$$
\ti r(i\om)=\fr1{\om}i({\cal L}_{12}B_v^{-1}+{\cal L}_{3})r_0+
\fr1{\om^2}{\cal L}_{12}(\vp_0+\Phi(i\om)+\Psi(i\om))
$$
$$
=\fr{{\cal L}_{12}}{\om}\left[i(B_v^{-1}+{\cal L}_{12}^{-1}{\cal L}_{3})r_0+
\fr{\vp_0+\Phi(i\om)+\Psi(i\om)}{\om}\right]=
\fr{{\cal L}_{12}}{\om}\left[i(B_v^{-1}+{\cal L}_{12}^{-1}{\cal L}_{3})r_0+
\ti g(\om)\right]
$$
$$
={\cal L}_{12}\fr{i(B_v^{-1}+{\cal L}_{12}^{-1}{\cal L}_{3})r_0+\ti g(0)+\ti g(\om)-
\ti g(0)}{\om}={\cal L}_{12}\fr{\ti g(\om)-\ti g(0)}{\om},
$$
since $i(B_v^{-1}+{\cal L}_{12}^{-1}{\cal L}_{3}(0))r_0+\ti g(0)=0$ by the
symplectic orthogonality conditions (\re{ort123}), because
$\ti g(0)=i(\Phi'(0)+\Psi'(0))$. Thus, $r_1(t)=\Lam^{-1}\zeta_1(\om){\cal L}_{12}*h(t)$
by (\re{QP1i3}), where
$$
h(t):=\Lam^{-1}\fr{\ti g(i\om)-\ti g(0)}{\om}=i\int\limits^{t}_{\infty}g(s)ds.
$$
This integral decays like $Ct^{-1-\de}\Vert X_0\Vert_\beta$ for $t\ge0$ by
(\re{fdec}), hence $|r_1(t)|\le Ct^{-1-\de}\Vert X_0\Vert_\beta$ for $t\ge0$.\hfill
$\Box$

Now Theorem \re{171} is proved.

%%%%%%%%%%%%%%%%%%%%%%%%%%%%%%%%%%%%%%%%%%%%%%%%%%%%%%%%%%
\setcounter{equation}{0}
\section{Time Decay of Fields}

Here we construct the field components $e(x,t),a(x,t)$ of the
solution $X(t)$
and prove their decay corresponding to (\re{frozenest}).
Let us denote
$
F(t)=(e(\cdot,t),a(\cdot,t)).
$
We will construct the fields solving the first two equations of
 (\re{AAA}), where $A$ is given by (\re{Api}).
These two equations have the form
\be\la{linf}
\dot F(t)=\left(
\ba{ll}
v\cdot\na & -\De \\
-1 & v\cdot\na
\ea
\right)F+\left(
\ba{l}
\Pi(t) \\
0
\ea
\right),\,\,\,\,\Pi(t):=\Pi_s(r(t)\na\rho v-\rho B_v\vp(t)).
\ee
By
Theorem \re{171}
 we know that $r(t)$ and $\vp(t)$ are continuous and
\be\la{linb}
|r(t)|+|\vp(t)|\le \ds\fr{ C(\rho, \ti v)\Vert X_0\Vert_{\beta} }
{ (1+t)^{1+\de} },~~~~
t\ge 0.
\ee
Hence, the Proposition \re{lindecay}
is reduced now to the following
\begin{pro} \la{pfi}
i) Let functions $r(t),\vp(t)\in C([0,\infty);\R^3)$,
and $F_0\in\cF$.
Then
the equation (\re{linf}) admits a unique
solution $F(t)\in C[0,\infty;\cF)$ with the initial condition
$F(0)=F_0$.
\\
ii) If $X_0=(F_0;r_0,\vp_0)\in{\cal E}_{\beta}$ and the decay (\re{linb}) holds,
the corresponding fields also decay
uniformly in $v$:
\be\la{lins}
\Vert F(t)\Vert_{-2-\de}\le \ds\fr{C(\rho,\ti v)\Vert X_0\Vert_{\beta}}{(1+t)^{1+\de}}
,~~~~t\ge 0,
\ee
for $|v|\le \ti v$ with any $\ti v\in (0;1)$.
\end{pro}
{\bf Proof }  Both statements follow from the Duhamel representation
\be\la{Duh} F(t)=W(t)F_0+\left[\int_0^tW(t-s)\left( \ba{l} \Pi(s) \\ 0 \ea \right)ds
\right],~~~~~~t\ge 0,
\ee
where $W(t)$ is the dynamical group of the modified wave equation
\be\la{184st} \dot F(t)=\left( \ba{cc} v\cdot\na & -\De \\ -1 & v\cdot\na \ea
\right)F(t)
\ee
and from the following decay properties of the group $W(t)$:

\bl\label{lemma}
For $\ti{v}<1$ and $F_{0}\in \mathcal{F}_{\al}$, $\al>1$, the following decay holds,
\begin{equation}
\Vert W(t)F_{0}\Vert _{-\al}\leq \frac{C(\al ,\ti{v})}{(1+t)^{\al-1} }
\Vert F_{0}\Vert _{\al },~~~~~~t\geq 0,  \label{dede}
\end{equation}
for the dynamical group $W(t)$ corresponding to the modified wave equation
(\ref{184st}) with $|v|<\ti{v}$.
\el

For details see proof of \ci[Lemma 10.2]{IKV09}.
Proposition \re{lindecay} is proved.

%%%%%%%%%%%%%%%%%%%%%%%%%%%%%%%%%%%%%%%%%%%%%%%%%%%%%%%%%
%%%%%%%%%%%%%%%%%%%%%%%%%%%%%%%%%%%%%%%%%%%%%%%%%%%%%%%%%

\appendix

\setcounter{equation}{0}

\section{Computing Symplectic Form}

Here we compute the matrix elements $\Om(\tau_j,\tau_l)$ of
the matrix $\Om$ and prove that the matrix is non-degenerate.
 For $j,l=1,2,3$ it follows from (\re{inb}) and (\re{OmJ})
that \be\la{jl} \Om(\tau_j,\tau_l) =\langle\pa_j E_v,\pa_l A_v\rangle-\langle\pa_j
A_v,\pa_l E_v\rangle, \,\,\,
\Om(\tau_{j+3},\tau_{l+3})=\langle\pa_{v_j}E_v,\pa_{v_l}A_v\rangle-
\langle\pa_{v_j}A_v,\pa_{v_l}E_v\rangle, \ee \be\la{Wx} \Om(\tau_{j},\tau_{l+3})
=-\langle\pa_{j}E_v,\pa_{v_l}A_v\rangle+
\langle\pa_{j}A_v,\pa_{v_l}E_v\rangle+e_j\cdot\pa_{v_l}P_v. \ee In Fourier
representation the solitons read \be\la{Sfur} \hat E_v(k)=\fr{i(kv)\hat\rho}{\hat
D_0}\left(\fr{(kv)}{k^2}k-v\right),\,\,\,\hat A_v(k)=\fr{-\hat\rho}{\hat
D_0}\left(\fr{(kv)}{k^2}k-v\right), \ee \be\la{Pvfur} P_v=p_v+\langle
A_v,\rho\rangle=p_v+v\int\fr{|\hat\rho|^2dk}{\hat
D_0}-\int\fr{|\hat\rho|^2dk}{k^2\hat D_0}(kv)k, \ee where $\hat D_0:=k^2-(kv)^2$;
$\hat D_0$ is nonnegative and even in $k$. Differentiating in $v$ we obtain for
$j=1,2,3$: \be\la{dervE} \!\!\!\!\!\!\!\!\! \pa_{v_j}\hat E_v\!=\!\fr{i\hat\rho}{\hat
D_0}\left(\fr{2k_j(kv)}{\hat D_0}k-\fr{k_j(k^2+(kv)^2)}{\hat D_0}v-(kv)e_j\right), \,
\pa_{v_l}\hat A_v\!=\!\fr{\hat\rho}{\hat D_0}\left(\fr{2k_l(kv)}{\hat
D_0}v-\fr{k_l(k^2+(kv)^2)}{k^2\hat D_0}k+e_l\right), \ee \be\la{dervP}
\pa_{v_l}P_v=\pa_{v_l}p_v+\langle\pa_{v_l}A_v,\rho\rangle=
B_v^{-1}e_l+\int\fr{|\hat\rho|^2dk}{\hat D_0}e_l+2\int\fr{|\hat\rho|^2(kv)k_ldk}{\hat
D_0^2}v-\int\fr{|\hat\rho|^2(k^2+(kv)^2)k_ldk}{k^2\hat D_0^2}k. \ee Then for
$j,l=1,2,3$ we get from (\re{jl}) by the  Parseval identity,
$$
\langle\pa_j E_v,\pa_l A_v\rangle=-i\int\fr{k_jk_l(kv)|\hat\rho|^2}{\hat D_0^2}
\left(\fr{(kv)}{k^2}k-v\right)^2dk=0,
$$
since the integrand function is odd in $k$. Similarly, $\langle\pa_j A_v,\pa_l
E_v\rangle=0$ and thus $\Om(\tau_j,\tau_l) = 0$. Further, by (\re{jl}),
$$
\langle\pa_{v_j}E_v,\pa_{v_l}A_v\rangle=i\int\fr{|\hat\rho|^2}{\hat D_0^2}
\left(\fr{4k_jk_l(kv)^3}{\hat D_0^2}+\fr{2k_jk_l(kv)}{\hat D_0}-\right.
$$
$$
\fr{2k_jk_l(kv)(k^2+(kv)^2)}{\hat D_0^2}-\fr{2k_jk_l(kv)(k^2+(kv)^2)v^2}{\hat D_0^2}
-\fr{k_j(k^2+(kv)^2)v_l}{\hat D_0}+
$$
$$
\left.\fr{k_jk_l(kv)(k^2+(kv)^2)^2}{k^2\hat D_0^2}-\fr{2k_l(kv)^2v_j}{\hat
D_0}-(kv)\de_{jl}+\fr{k_jk_l(kv)(k^2+(kv)^2)}{k^2\hat D_0}\right)dk=0,
$$
since the integrand function is odd in $k$. Note that the integral converges by the
neutrality condition (\re{neutr}). Similarly,
$\langle\pa_{v_j}A_v,\pa_{v_l}E_v\rangle$=0 and thus, $\Om(\tau_{j+3},\tau_{l+3})=0$.
 Now let us compute $\Om(\tau_{j},\tau_{l+3})$. First,
$$
-\langle\pa_j E_v,\pa_{v_l}A_v\rangle=\int\fr{k_j(kv)|\hat\rho|^2dk}{\hat D_0^2}
\left(v_l+\fr{2k_l(kv)v^2}{\hat D_0}-\fr{k_l(kv)(k^2+(kv)^2)}{k^2\hat D_0}\right).
$$
Second,
$$
\langle\pa_j A_v,\pa_{v_l}E_v\rangle=\int\fr{k_j|\hat\rho|^2dk}{\hat D_0^2}
\left((kv)v_l+\fr{k_l(k^2+(kv)^2)v^2}{\hat D_0}-\fr{2k_l(kv)^2}{\hat D_0}\right).
$$
And third,
$$
e_j\cdot\pa_{v_l}P_v=e_j\cdot B_v^{-1}e_l+\int\fr{|\hat\rho|^2dk}{\hat D_0}\de_{jl}
+2\int\fr{|\hat\rho|^2(kv)k_ldk}{\hat
D_0^2}v_j-\int\fr{|\hat\rho|^2(k^2+(kv)^2)k_jk_ldk}{k^2\hat D_0^2}.
$$
By a straightforward computation we obtain that the matrix
$\Om^+(v)=\Vert\Om(\tau_j,\tau_{l+3})\Vert\vert_{j,l=1,2,3}$ is positive definite and
hence non-degenerate. Finally, the matrix \be\la{Omega}
\Vert\Om(\tau_j,\tau_{l})\Vert\vert_{j,l=1,...,6}= \left( \ba{ll}
0 & \Om^+(v) \\
-\Om^+(v) & 0 \ea \right) \ee is also non-degenerate.

%%%%%%%%%%%%%%%%%%%%%%%%%%%%%%%%%%%%%%%%%%%%%%%%%%%%%%%%%%%
%%%%%%%%%%%%%%%%%%%%%%%%%%%%%%%%%%%%%%%%%%%%%%%%%%%%%%%%%%%

\setcounter{equation}{0}

\section{Bounds for the matrix $M^{-1}(i\om)$}

{\bf Proposition B.1}
{\it The following bound holds,} \be\la{Minvest} \Vert
M^{-1}(i\om)\Vert={\cal O}\left(\fr1{|\om|}\right),\,\,\,\om\to\infty.
\ee
\pru Recall that
$$
M(i\om)=\left( \ba{ll}
i\om E & -B_v \\
C(i\om)+G & i\om E-F(i\om) \ea \right),
$$
where
$$
B_v=\left( \ba{lll}
\nu^3 & 0 & 0 \\
0 & \nu & 0 \\
0 & 0 & \nu
\ea \right),\,\,\,C(i\om)=\left( \ba{lll}
c_1(i\om) & 0 & 0 \\
0 & c(i\om) & 0 \\
0 & 0 & c(i\om)
\ea \right),
$$
$$
G=\left( \ba{lll}
g_1 & 0 & 0 \\
0 & g & 0 \\
0 & 0 & g
\ea \right),\,\,\,F(i\om)=\left( \ba{lll}
f_1(i\om) & 0 & 0 \\
0 & f(i\om) & 0 \\
0 & 0 & f(i\om)
\ea \right).
$$
Here $\nu=\sqrt{1-v^2}$ and $c_1, c; g_1, g; f_1, f$ are defined by (\re{CG}),
(\re{Cdet}); (\re{G}), (\re{gg23}); (\re{Fdet}) respectively. Further,
\be\la{detMD1D}
\det~M(i\om)=d_1d^2,\,\,\,d_1=i\om(i\om-f_1)+\nu^3(c_1+g_1),\,\,\,d=i\om(i\om-f)+
\nu(c+g).
\ee

{\bf Lemma B.2} {\it The functions $c_1(i\om), c(i\om), f_1(i\om), f(i\om)$ are
bounded for   $\om\in\R$ with large $|\om|$.}

\pru Let us consider only the first function $c_1(i\om)$ in detail, for the rest
three functions the argument is similar. By (\re{CG})
$$
c_1(i\om)=i|v|\int\,k_1\fr{(i\om+ik_1|v|)|\hat\rho|^2}{\hat D(i\om)}
(1-\fr{k_1^2}{k^2})dk=
$$
$$
i|v|\int\,k_1\fr{(i\om+ik_1|v|)|\hat\rho|^2}{\hat
D(i\om)}dk-i|v|\int\,k_1\fr{(i\om+ik_1|v|)|\hat\rho|^2}{\hat D(i\om)}
\fr{k_1^2}{k^2}dk=:{\cal C}_1(i\om)+{\cal C}_2(i\om).
$$
Let us study ${\cal C}_1(i\om)$, for ${\cal C}_2(i\om)$ the argument
is similar. In
the $x$-space we obtain:
\be\la{C1x} {\cal C}_1(i\om)=|v|\langle
-\pa_1(i\om-|v|\pa_1)\rho,D^{-1}(i\om)\rho\rangle.
\ee
where $D(i\om)=-\De+(i\om-v\na)^2$ is the differential operator
with the symbol (\re{Dlam}).
Hence, ${\cal C}_1(i\om)$ is bounded by the decay
\be\la{agm}
|\langle\pa_1^2\rho,D^{-1}(i\om)\rho\rangle|={\cal O}(\fr1{|\om|}),
~~~~~~~~~~
|\om|\to\infty
\ee
which follows
from the
Agmon estimates \ci[(A.2')]{Ag} (see also Appendix in \ci{KK10}).
The Agmon  estimates are applicable to the operator
$D(i\om)$ because of
representation
\be\la{Dlamr}
D(i\om)=e^{-i\ga v_1x_1}[-(1-v_1^2)\pa_1^2-\pa_2^2-\pa_3^2-\om^2]
e^{i\ga v_1x_1}
\ee
with $(1-v_1^2)\ga =\om$ in the
coordinates where $v=(v_1,0,0)$.
\hfill$\Box$

{\bf Corollary B.3} The determinants
{\it $d_1(i\om)$ and $d(i\om)$ are nonzero
for $\om\in\R$ with large $|\om|$.
}

\bigskip

Further, the inverse matrix reads \be\la{Minvlam} M^{-1}(i\om)=\left( \ba{ll}
L_{11} & L_{12} \\
L_{21} & L_{22} \ea \right), \ee where \be\la{L1112} L_{11}(i\om)=\left( \ba{lll}
(i\om-f_1)/d_1 & 0 & 0 \\
0 & (i\om-f)/d & 0 \\
0 & 0 & (i\om-f)/d \ea \right),\,\,\,L_{12}(i\om)=\left( \ba{lll}
\nu^3/d_1 & 0 & 0 \\
0 & \nu/d & 0 \\
0 & 0 & \nu/d \ea \right),
\ee
\be\la{L2122}
L_{21}(i\om)=\left( \ba{lll}
-(c_1+g_1)/d_1 & 0 & 0 \\
0 & -(c+g)/d & 0 \\
0 & 0 & -(c+g)/d \ea \right),\,\,\,L_{22}(i\om)=\left( \ba{lll}
i\om/d_1 & 0 & 0 \\
0 & i\om/d & 0 \\
0 & 0 & i\om/d \ea \right).
\ee
Now Proposition B.1 follows from
(\re{detMD1D}), (\re{Minvlam})--(\re{L2122}), and Lemma B.2.\hfill$\Box$

\subsection{Proof of Lemma \re{PiomQiom}}

By Lemma \re{cac} the functions $c_{1}(\lam),c(\lam);f_{1}(\lam),f(\lam)$ are
analytic in $\C$. Thus,
$$
c_1(\lam)=c_1(0)+c_1'(0)\lam+\fr{c_1''(0)}{2}\lam^2+\dots,\,\,\,
c(\lam)=c(0)+c'(0)\lam+\fr{c''(0)}{2}\lam^2+\dots,
$$
$$
f_1(\lam)=f_1(0)+f_1'(0)\lam+\dots,\,\,\,f(\lam)=f(0)+f'(0)\lam+\dots.
$$
Below we write $v$ instead of $|v|$ for simplicity of notations. By (\re{CG}),
(\re{Fdet}), (\re{f123})
$$
c_1'(\lam)=iv\int dk|\hat\rho|^2k_1\left(1-\fr{k_1^2}{k^2}\right)\fr{k^2-
(\lam+ik_1v)^2}{\hat D^2(\lam)},
$$
\be\la{f1p} f_1'(\lam)=\nu^3\int
dk|\hat\rho|^2\left(\fr{k_1^2}{k^2}-1\right)\fr{k^2-(\lam+ik_1v)^2}{\hat D^2(\lam)}.
\ee
Further, $c'(\lam)=(c_{12})'(\lam)+(c_{22})'(\lam)$,
$f'(\lam)=(f_{12})'(\lam)+(f_{22})'(\lam)$. By (\re{f123}), (\re{CG}),

(\re{c2f2lam})
$$
(c_{12})'(\lam)=-\fr{iv}2\int
dk|\hat\rho|^2k_1(1-\fr{k_1^2}{k^2})\fr{k^2-(\lam+ik_1v)^2}{\hat
D^2(\lam)},
\,\,\,
(c_{22})'(\lam)=v^2\int dk|\hat\rho|^2\fr{(k^2-k_1^2)(\lam+ik_1v)}
{\hat D^2(\lam)},
$$
\be\la{f1plam} (f_{12})'(\lam)=-\fr{\nu}2\int
dk|\hat\rho|^2\left(1+\fr{k_1^2}{k^2}\right)\fr{k^2-(\lam+ik_1v)^2}{\hat
D^2(\lam)},\,\,\,(f_{22})'(\lam)=-2i\nu v\int dk|\hat\rho|^2k_1
\fr{(\lam+ik_1v)}{\hat
D^2(\lam)}.
\ee
Further, \be\la{c1pplam} c_1''(\lam)=-2iv\int
dk|\hat\rho|^2k_1\left(1-\fr{k_1^2}{k^2}\right)\fr{(\lam+ik_1v)(3k^2-
(\lam+ik_1v)^2)}{\hat
D^3(\lam)}, \ee and $c''(\lam)=(c_{12})''(\lam)+(c_{22})''(\lam)$,
where \be\la{c1pp}
(c_{12})''(\lam)=iv\int
dk|\hat\rho|^2k_1(1-\fr{k_1^2}{k^2})\fr{(\lam+ik_1v)(3k^2-(\lam+ik_1v)^2)}
{\hat
D^3(\lam)}, \ee \be\la{c2pp} (c_{22})''(\lam)=v^2\int
dk|\hat\rho|^2\fr{(k^2-k_1^2)(k^2-3(\lam+ik_1v)^2)}{\hat
D^3(\lam)}.
\ee
Note that
$c_1(0)=-g_1$, $c(0)=-g$, $c_1'(0)=c'(0)=0$, then
$$
c_1(\lam)=-g_1+\lam^2I_1(\lam),\,\,\,c(\lam)=-g+\lam^2I(\lam),
$$
where the functions $I_1(\lam)$, $I(\lam)$ are analytic in $\C$ and $I_1(0)=
c_1''(0)/2$, $I(0)=c''(0)/2$. By (\re{c1pplam}) to (\re{c2pp}) we have
\be\la{c1pp0}
c_1''(0)=2v^2\int dk|\hat\rho|^2k_1^2\left(1-\fr{k_1^2}{k^2}\right)\fr{3k^2+
(k_1v)^2}{(k^2-(k_1v)^2)^3},
\ee
\be\la{cpp0}
c''(0)=v^2\int dk|\hat\rho|^2\fr{(k^2-k_1^2)(k^2(k^2+3(k_1v)^2)-k_1^2(3k^2+
(k_1v)^2))}{k^2(k^2-(k_1v)^2)^3}.
\ee
Similarly, $f_1(0)=f(0)=0$ and
$$
f_1(\lam)=\lam J_1(\lam),\,\,\,f(\lam)=\lam J(\lam),
$$
where the functions $J_1(\lam)$, $J(\lam)$ are analytic in $\C$ and $J_1(0)=f_1'(0)$,
$J(0)=f'(0)$. By (\re{f1plam}) we have
\be\la{f1p0}
f_1'(0)=\nu^3\int dk|\hat\rho|^2(\fr{k_1^2}{k^2}-1)\fr{k^2+(k_1v)^2}{(k^2-(k_1v)^2)^2},
\ee
\be\la{fp0}
f'(0)=\fr{\nu}2\int dk|\hat\rho|^2\fr{(3k^2(k_1v)^2-k_1^2(k_1v)^2-k^2(k^2+k_1^2))}
{k^2(k^2-(k_1v)^2)^2}.
\ee
We put $\lam=i\om$ and obtain
\be\la{cgf}
c_1(i\om)+g_1=-\om^2I_1(i\om),\,\,c(i\om)+g=-\om^2I(i\om),\,\,f_1(i\om)=
i\om J_1(i\om),\,\,f(i\om)=i\om J(i\om).
\ee
Then
\be\la{D1D}
d_1(i\om)=-\om^2(1-J_1(i\om)+\nu^3I_1(i\om)),\,\,\,d(i\om)=
-\om^2(1-J(i\om)+\nu I(i\om))
\ee
Substitute (\re{cgf}), (\re{D1D}) to (\re{L1112}), (\re{L2122}) and obtain
$$
L_{11}(i\om)=\fr1{\om}{\cal L}_{11}(i\om),\,\,L_{12}(i\om)=
\fr1{\om^2}{\cal L}_{12}(i\om),\,\,L_{21}(i\om)={\cal L}_{21}(i\om),\,\,L_{22}(i\om)=
\fr1{\om}{\cal L}_{22}(i\om),
$$
where (we omit the dependance on $i\om$ for simplicity of notations)
$$
{\cal L}_{11}=\left( \ba{lll}
\fr{\ts i(J_1-1)}{\ts 1-J_1+\nu^3I_1} & 0 & 0 \\
0 & \fr{\ts i(J-1)}{\ts 1-J+\nu I} & 0 \\
0 & 0 & \fr{\ts i(J-1)}{\ts 1-J+\nu I} \ea\right),
$$
$$
{\cal L}_{12}=\left( \ba{lll}
\fr{\ts-\nu^3}{\ts 1-J_1+\nu^3I_1} & 0 & 0 \\
0 & \fr{\ts-\nu}{\ts 1-J+\nu I} & 0 \\
0 & 0 & \fr{\ts-\nu}{\ts 1-J+\nu I} \ea\right),
$$
$$
{\cal L}_{21}=\left( \ba{lll}
\fr{\ts -I_1}{\ts 1-J_1+\nu^3I_1} & 0 & 0 \\
0 & \fr{\ts -I}{\ts 1-J+\nu I} & 0 \\
0 & 0 & \fr{\ts -I}{\ts 1-J+\nu I} \ea\right),
$$
$$
{\cal L}_{22}=\left( \ba{lll}
\fr{\ts -i}{\ts 1-J_1+\nu^3I_1} & 0 & 0 \\
0 & \fr{\ts -i}{\ts 1-J+\nu I} & 0 \\
0 & 0 & \fr{\ts -i}{\ts 1-J+\nu I} \ea\right).
$$
Note that \be\la{M1112} {\cal L}_{11}(\om)=i{\cal L}_{12}(\om)B_v^{-1}+i{\cal L}_3, \ee
where
\be\la{calL3}
{\cal L}_3=\left(
\ba{ccc}
\fr{\ts J_1}{\ts 1-J_1+\nu^3I_1} & 0 & 0 \\
0 & \fr{\ts J}{\ts 1-J+\nu I_1} & 0 \\
0 & 0 & \fr{\ts J}{\ts 1-J+\nu I_1}
\ea
\right).
\ee
Finaly, we prove that the denominators of the matrix
elements of each matrix ${\cal L}_{11}$ to ${\cal L}_{22}$ and ${\cal L}_3$ are
nonzero at $\om=0$. Indeed, $-J_1(0)+\nu^3I_1(0)>0$, since $I_1(0)>0$ and $J_1(0)<0$
by (\re{c1pp0}), (\re{f1p0}). Further, by a straightforward computation we obtain that
$$
-J(0)+\nu I(0)=\fr{\nu}2\int\,dk\fr{|\hat\rho|^2(k^6(1+v^2)+k^4k_1^2(1+3v^4-8v^2)+
k^2k_1^4v^2(3-v^2))}{k^2(k^2-(k_1v)^2)^3}.
$$
It is easy to check that $k^6(1+v^2)+k^4k_1^2(1+3v^4-8v^2)+k^2k_1^4v^2(3-v^2)\ge0$.
This completes the proof of Lemma \re{PiomQiom}.\hfill$\Box$

%%%%%%%%%%%%%%%%%%%%%%%%%%%%%%%%%%%%%%%%%%%%%%%%%%%%%%%%%%%%%%%%
%%%%%%%%%%%%%%%%%%%%%%%%%%%%%%%%%%%%%%%%%%%%%%%%%%%%%%%%%%%%%%%%

\setcounter{equation}{0}

\section{Symplectic Orthogonality Conditions}

For $j=1,2,3$ we have
$$
0=\Om(Z_0,\tau_j)=-\langle e_0,\pa_j A_v\rangle+\langle a_0,\pa_j
E_v\rangle-(\vp_0+\langle a_0,\rho\rangle)\cdot e_j=
$$
$$
-\int dk\,\,\hat e_0\,\,\overline{\fr{-ik_j(-\hat\rho)}{\hat
D}(\fr{kv}{k^2}k-v)}+\int dk\,\,\hat a_0\,\,\overline{\fr{-ik_ji(kv)\hat\rho}{\hat
D}(\fr{kv}{k^2}k-v)}-(\vp+\langle a_0,\rho\rangle)\cdot e_j.
$$
Since $\hat e_0\bot k$ and $\hat a_0\bot k$, the condition simplifies to
$$
-i\int dk\,\,\hat e_0\,\,\fr{k_j\ov{\hat\rho}}{\hat D}v-\int dk\,\,\hat
a_0\,\,\fr{k_j(kv)\ov{\hat\rho}}{\hat D}v-\int dk\hat a_0\ov{\hat\rho}\cdot
e_j-\vp_0\cdot e_j=0,
$$
or, in the vector form,
\be\la{ort123C}
\vp_0+\int dk\fr{\ov{\hat\rho}}{\hat D}[i(\hat e_0v)k+(\hat a_0v)(kv)k+k^2\hat a_0-
(kv)^2\hat a_0]=0.
\ee
On the other hand,
\be\la{phiPhi}
\vp_0+\Phi(0)=\vp_0+\int dk\fr{\ov{\hat\rho}}{\hat D}[i(kv)\hat e_0+k^2\hat a_0].
\ee
Subtract (\re{ort123C}) from (\re{phiPhi}) and obtain
$$
\vp_0+\Phi(0)=-i\int dk\fr{\ov{\hat\rho}}{\hat D}[(\hat e_0v)k-(kv)\hat e_0]-
\int dk\fr{(kv)\ov{\hat\rho}}{\hat D}[(\hat a_0v)k-(kv)\hat a_0]=
$$
$$
\int dk\fr{\ov{\hat\rho}}{\hat D}v\we((-ik)\we\hat e_0)+\int dk\fr{\ov{\hat\rho}}
{\hat D}(-ikv)v\we((-ik)\we\hat a_0)=-\Psi(0).
$$
Thus, (\re{ort123C}) reads $\vp_0+\Phi(0)+\Psi(0)=0$.

\bigskip

\noindent Further,
the symplectic orthogonality conditions $\Om(Z_0,\tau_{j+3})$, $j=1,2,3$ in the vector
form read
$$
0=\int dk\fr{\ov{\hat\rho}}{\hat D}\left[\fr{2(kv)(\hat e_0v)}
{\hat D}k+\hat e_0\right]-i\int dk\fr{\ov{\hat\rho}}{\hat D}\left[\fr{k^2+(kv)^2}
{\hat D}(\hat a_0v)k+(kv)\hat a_0\right]+
$$
\be\la{ort456}
B_v^{-1}r_0+\int dk\fr{|\hat\rho|^2}{\hat D}r_0+2\int dk\fr{|\hat\rho|^2(kv)(r_0v)}
{\hat D^2}k
-\int dk\fr{|\hat\rho|^2(k^2+(kv)^2)(r_0k)}{k^2\hat D^2}k.
\ee
The second line involving $r_0$, in the coordinate system, where $v=(v,0,0)$,
simplifies to
$\,\,\,B_v^{-1}r_0+$
$$
\left(
\ba{ccc}
\int\,dk\fr{|\hat\rho|^2((k^2-k_1^2)(k^2+(k_v)^2))}{k^2D^2} & 0 & 0\\
0 & \int\,dk\fr{|\hat\rho|^2(k^2(k^2+k_1^2)+k_1^2(k_v)^2-3k^2(k_1v)^2)}{2k^2D^2} & 0 \\
0 & 0 & \int\,dk\fr{|\hat\rho|^2(k^2(k^2+k_1^2)+k_1^2(k_v)^2-3k^2(k_1v)^2)}{2k^2D^2}
\ea
\right)r_0.
$$
And this is exactly $(B_v^{-1}+{\cal L}_{12}{\cal L}_{3}(0))r_0$, since
$$
{\cal L}_{12}{\cal L}_{3}(0)=B_v^{-1}\left(
\ba{ccc}
-J_1(0) & 0 & 0 \\
0 & -J(0) & 0 \\
0 & 0 & -J(0)
\ea
\right),
$$
where $J_1(0)=f_1'(0)$, $J(0)=f'(0)$, see (\re{f1p0}), (\re{fp0}).

Finally, by (\re{Phioflam}), (\re{Psilam})
$$
\Phi'(0)+\Psi'(0)=
$$
$$
\int\,dk\fr{\ov{\hat\rho}[D\hat e_0-i(kv)D\hat a_0+2(kv)(v\cdot\hat e_0)k-
i(k^2+(kv)^2)(v\cdot\hat a_0)k]}{D^2}
$$
wich coincides with the first line of (\re{ort456}) involving $\hat e_0, \hat a_0$.

%%%%%%%%%%%%%%%%%%%%%%%%%%%%%%%%%%%%%%%%%%%%%%%%%%%%%%%%%%%%

\end{document}